\documentclass[reprint,twocolumn,aps,prl,nopacs,superscriptaddress]{revtex4-2}
\usepackage{float}
\usepackage[english]{babel}
\usepackage[utf8]{inputenc}
\usepackage{fancyhdr}
\usepackage{sidecap}
\usepackage{multirow}
\usepackage{tabularx}

\usepackage{graphicx}
\usepackage{braket}
\usepackage{tikz}
\usepackage{color}
\usepackage{amsmath}
\usepackage[colorlinks=True, linkcolor=blue, filecolor=magenta, urlcolor=blue,citecolor=blue]{hyperref}

\usepackage{graphicx}
\usepackage{comment}
\usepackage{multirow}
\usepackage{xspace}

\newcommand{\subm}[1]{_{\mathrm {#1}}}
\usepackage{xcolor}
\newcommand{\SRO}{Sr$_2$RuO$_4$\xspace}

\begin{document}
%%%%%%%%% Title %%%%%%%%%
\title{\Large{\textbf {Muon Knight shift as a precise probe of the superconducting symmetry of Sr$_2$RuO$_4$}}}
%%%%%%%%% Authors name %%%%%
\author{Hisakazu Matsuki}
 \email{matsuki.hisakazu.2i@kyoto-u.ac.jp}
\affiliation{Toyota Riken–Kyoto University Research Center (TRiKUC), Kyoto University, Kyoto 606-8501, Japan}
\affiliation{Institute for Chemical Research, Kyoto University, Uji, Kyoto 611-0011, Japan}

\author{Rustem Khasanov}
 \email{rustem.khasanov@psi.ch}
\affiliation{PSI Center for Neutron and Muon Sciences CNM, 5232 Villigen PSI, Switzerland}

\author{Jonas A. Krieger}
\affiliation{PSI Center for Neutron and Muon Sciences CNM, 5232 Villigen PSI, Switzerland}

\author{Thomas J. Hicken}
\affiliation{PSI Center for Neutron and Muon Sciences CNM, 5232 Villigen PSI, Switzerland}

\author{Kosuke Yuchi}
\affiliation{Institute for Solid State Physics, the University of Tokyo, Kashiwa, Chiba 277-8501, Japan}

\author{Jake S. Bobowski}
\affiliation{Department of Physics, University of British Columbia, Kelowna, BC V1V 1V7, Canada}

\author{Giordano Mattoni}
\affiliation{Toyota Riken–Kyoto University Research Center (TRiKUC), Kyoto University, Kyoto 606-8501, Japan}

\author{Atsutoshi Ikeda}
\affiliation{Toyota Riken–Kyoto University Research Center (TRiKUC), Kyoto University, Kyoto 606-8501, Japan}
\affiliation{Department of Electronic Science and Engineering, Graduate School of Engineering, Kyoto University, Kyoto 615-8510, Japan}

\author{Ryutaro Okuma}
\affiliation{Institute for Solid State Physics, the University of Tokyo, Kashiwa, Chiba 277-8501, Japan}

\author{Hubertus Luetkens}
\affiliation{PSI Center for Neutron and Muon Sciences CNM, 5232 Villigen PSI, Switzerland}

\author{Yoshiteru Maeno}
 \email{maeno.yoshiteru.b04@kyoto-u.jp}
\affiliation{Toyota Riken–Kyoto University Research Center (TRiKUC), Kyoto University, Kyoto 606-8501, Japan}

%%%%%%%%%%%%  abstract  %%%%%%%%%%%%
\begin{abstract}
Muon spin rotation ($\mu$SR) measurements of internal magnetic field shifts, known as the muon Knight shift, is used for determining pairing symmetries in superconductors. 
While this technique has been especially effective for $f$-electron-based heavy-fermion superconductors, it remains challenging in $d$-electron-based superconductors such as \SRO, where the Knight shift is intrinsically small. 
Here, we report high-precision muon Knight shift measurements of superconducting \SRO. 
We observe that using multiple pieces of crystals, a common practice in $\mu$SR measurements, induces a substantial paramagnetic shift below the superconducting transition temperature, $T\subm{c}$, when a weak magnetic field is applied. 
We attribute such an unresolved paramagnetic shift to stray fields generated by neighboring diamagnetic crystals. 
To avoid this, one piece of crystal was used in this study. We experimentally determine the muon Knight shift of \SRO in the normal state to be -116$\pm$7 ppm. 
By combining the observed muon Knight shift with independently determined bulk magnetization data from the same crystal used in $\mu$SR and carefully separating various contributions to the shift, we confirm a significant reduction in the spin Knight shift below $T\subm{c}$, consistent with spin-singlet-like pairing. 
This result constitutes the precise muon Knight shift measurement in a $d$-electron-based superconductor. Our results highlight the potential of $\mu$SR as a powerful complementary technique to the established method of nuclear magnetic resonance for probing the spin susceptibility in superconductors.
\end{abstract}
\maketitle

%%%%%%%%%%%  Introduction  %%%%%%%%%%%
The nature of a superconducting state is determined by the symmetry of its order-parameter, which dictates both the anisotropy of the energy gap in reciprocal $k$-space and the spin state of Cooper pairs. 
Despite significant advances in experimental and theoretical studies, identifying the precise order-parameter symmetry in unconventional superconductors--especially those with strong electron correlations--remains challenging.
A notable example is strontium ruthenate \SRO, where the order-parameter symmetry remains an open question after three decades of extensive study~\cite{maeno_superconductivity_1994,maeno_thirty_2024,maeno_still_2024}.

Since bulk magnetization of a superconductor is dominated by Meissner screening, microscopic techniques are required to extract its spin state.
A widely used approach is nuclear magnetic resonance (NMR), in which changes in the Knight shift--i.e., shifts in resonance frequency under an external magnetic field--probe the spin susceptibility below $T\subm{c}$~\cite{knight_nuclear_1956}. 
However, in highly conductive superconductors such as \SRO, eddy-current heating induced by NMR pulses may pose a serious issue~\cite{pustogow_constraints_2019, ishida_reduction_2020, chronister_evidence_2021}, making it essential to cross-check NMR results with alternative techniques.
Polarized neutron scattering provides one such alternative, though it typically requires long data acquisition times to yield sufficient statistical precision, restricting systematic field-dependent measurements~\cite{petsch_reduction_2020}.
In contrast, $\mu$SR extracts the Knight shift from muon spin precession frequency, proportional to the local magnetic field sensed by the muon~\cite{amato_introduction_2024,higemoto_spin_2016}. 
Its high frequency resolution and spin-1/2 simplicity make it a powerful probe of spin susceptibility in superconductors.

For many years, $\mu$SR studies of superconductors have been largely limited to $f$-electron-based heavy-fermion systems, owing to their large Knight shift signals in the normal state~\cite{heffner_muon_1986,uemura_muon_1986,luke_positive_1991,feyerherm_sr_1994,sonier_anomalous_2000,koda_evidence_2002,sonier_superconductivity_2003,higemoto_possible_2006,higemoto_spin-triplet_2007,higemoto_muon_2010,azari__2023,azari_coexistence_2025}. 
In contrast, spin-singlet $d$-electron-based superconductors exhibit intrinsically small muon Knight shifts, and unresolved paramagnetic shifts below $T\subm{c}$ have been reported~\cite{das_time-reversal_2020,graham_microscopic_2024}.
In those superconductors including \SRO,
the dipolar and spin-contact components can be comparable in magnitude but opposite in sign, further suppressing the overall signal. 
As a result, measuring the muon Knight shift in such systems is highly challenging. 
Consequently, past $\mu$SR studies of $d$-electron-based superconductors, including \SRO, 
have focused mainly on vortex-lattice structures~\cite{aegerter_evidence_1998,maisuradze_comparison_2009,ray_muon-spin_2014,yakovlev_atypical_2025}, magnetic penetration depth~\cite{ikeda_penetration_2020,khasanov_-plane_2023}, time-reversal-symmetry breaking (TRSB)~\cite{luke_time-reversal_1998,luke_unconventional_2000,shiroka_sr_2012,grinenko_unsplit_2021,grinenko_split_2021,grinenko_sr_2023,oudah_time-reversal_2024}, Hebel-Slichter peak below $T\subm{c}$~\cite{kiefl_coherence_1993,macfarlane_hebel-slichter_1994}, and magnetic ordering~\cite{nishida_first_1987,pumpin_internal_1988,brewer_antiferromagnetism_1988,higemoto_investigation_2014,fittipaldi_unveiling_2021}. 
A comparative summary of experimental techniques for Knight shift measurements in superconducting \SRO is discussed in the Supplemental Material (SM)~\cite{supplement_2025}, which includes Ref.~\cite{petsch_reduction_2020,pustogow_constraints_2019,ishida_reduction_2020,chronister_evidence_2021,jang_phase-locked_2011,jang_observation_2011,yonezawa_compact_2015,tenya_unusual_2006,kittaka_sharp_2014,shimizu_spin_2021,shull_neutron-diffraction_1966,mukuda_spin_1999,amato_introduction_2024,khasanov_-plane_2023,luke_time-reversal_1998,ghosh_recent_2020,xia_high_2006,xia_modified_2006,higemoto_spin_2016,bobowski_improved_2019,sato_simple_1989,huddart_intrinsic_2021,blundell_dft_2023,maeno_thirty_2024,hotz_experimental_2023,kittaka_angular_2009,maeno_two-dimensional_1997}. 
Recently, significant improvements in muon Knight shift measurements at sub-Kelvin temperatures enabled separate detection of the sample and Ag reference, which is used to calibrate the external magnetic field down to 0.05 K. 
Since the spectra now focus on the sample signal without overlapping with the Ag peak, the precision for detecting small frequency shifts is significantly improved.

\begin{figure}[t]
    \begin{center}
    \includegraphics[scale=0.26]{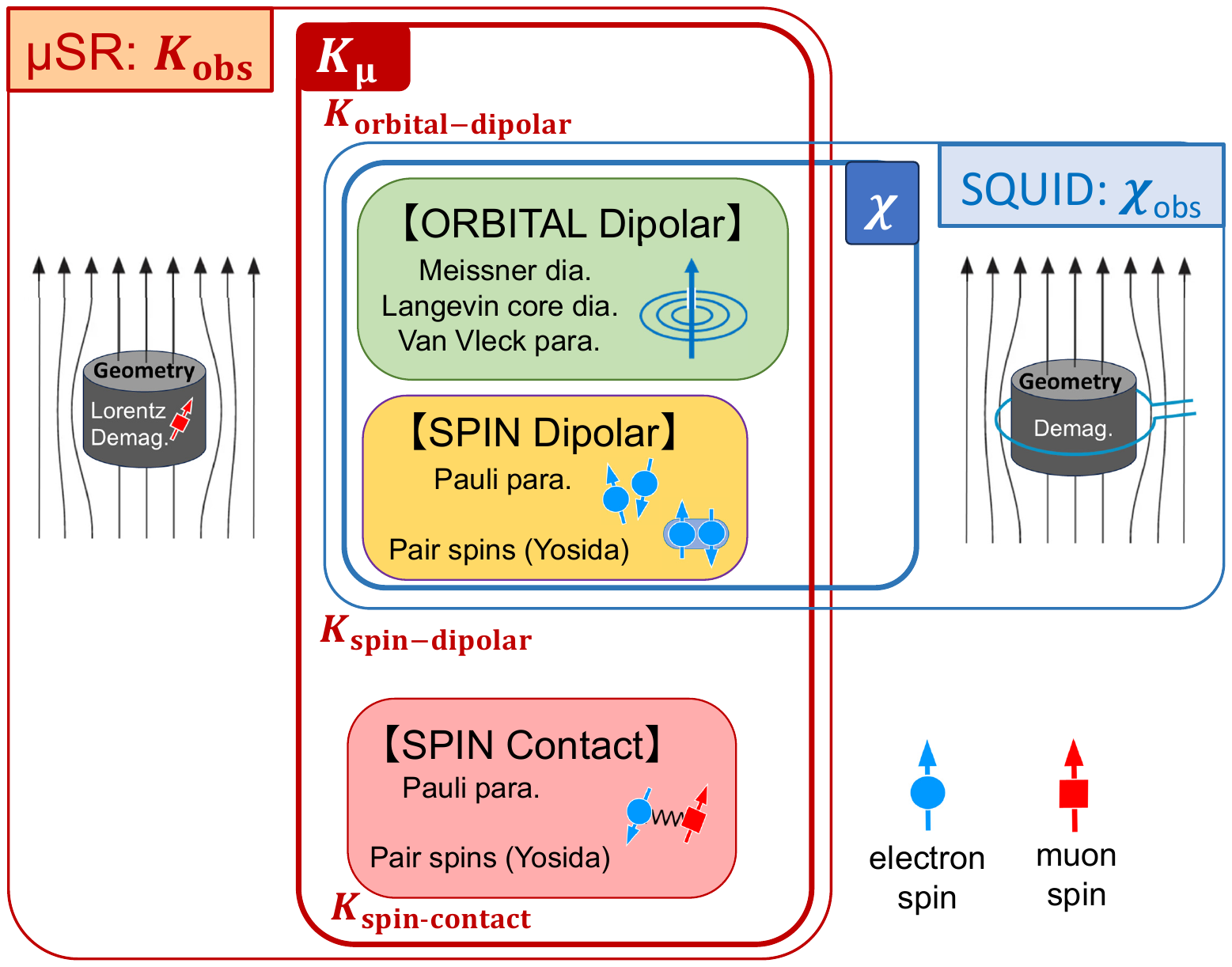}
    \caption{Schematic illustration of the contributions to the muon Knight shift $K\subm{obs}$ (thin red loop) and DC magnetic susceptibility measured by SQUID $\chi\subm{obs}$ (thin blue loop). 
    After correcting for geometry effects in both SQUID and $\mu$SR measurements, the remaining difference between the magnetic susceptibility $\chi$ (thick blue loop) probed by SQUID and the intrinsic muon Knight shift $K\subm{\mu}$ (thick red loop) probed by $\mu$SR corresponds to the spin contact term $K\subm{spin\text{-}contact}$, which directly reflects superconducting pairing symmetry.
    }
    \label{fig1}
    \end{center}
\end{figure}

In this $Letter$, we investigate the pairing symmetry of superconducting \SRO using $\mu$SR. We first highlight that measurements on multiple crystal pieces introduce stray fields, leading to spurious Knight shift artifacts.
To address this issue, we compare Knight shifts for one- and six-crystal configurations under identical conditions. Only the one-crystal measurement exhibits the expected diamagnetic shift below $T\subm{c}$, consistent with intrinsic superconducting behavior.
By combining $\mu$SR with bulk magnetization measurements on the same sample and separating field contributions as illustrated in Fig.~\ref{fig1}, we observe strong suppression of the spin Knight shift below $T\subm{c}$, a signature of spin-singlet-like pairing.\\

\begin{figure*}[t]
    \begin{center}
    \includegraphics[scale=0.96]{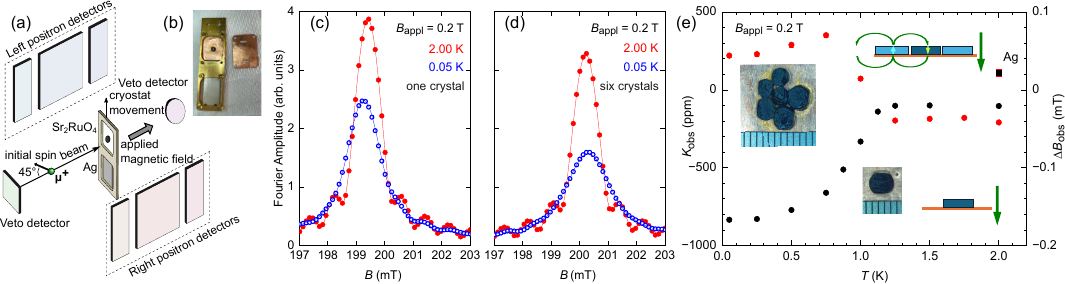}
    \caption{Schematic diagrams of (a) FLAME and (b) the tandem sample holder used for $\mu$SR. 
    (a) FLAME consists of a pair of three muon (positron) detectors (left and right) plus forward and backward veto detectors. The muon detectors collect positron signals, and the veto detectors catch muons missing the sample. 
    (b) The upper holder with one \SRO crystal on a 25-$\mu$m-thick copper-foil and the lower holder with a reference Ag plate. The upper holder is capped with another copper-foil, shown on the right, ensuring good thermal anchoring. 
    (c-d) The Fourier transforms of the muon time spectra at 2~K (red) and 0.05~K (blue) shown in SM~\cite{supplement_2025}. 
    The line-shape remains symmetric within the fitting range.
    (c) One-crystal, (d) Six-crystals.
    (e) Muon Knight shift $K\subm{obs}$ (left axis) and observed field shift $\Delta B\subm{obs}$ vs. temperature for two distinct sample configurations. A magnetic field of 0.2 T is applied along [100] direction, indicated by green arrows.
    Red circles: six-crystals (top left); black circles: one-crystal(bottom); scale bars in photos are 1~mm; red and black squares at 2~K: Knight shift of the Ag reference.
    The anticipated diamagnetic shift appears only for one-crystal, whereas measurements with six-crystals show an increase in $K\subm{obs}$ of 1100~ppm ($\approx$0.22~mT) at 0.05~K, attributable to additional stray-fields from neighboring diamagnetic crystals, as shown in the inset. 
    A smaller difference in $K\subm{obs}$ above $T\subm{c}$ also arises from stray-fields of neighboring paramagnetic crystals.}
    \label{fig2}
    \end{center}
\end{figure*}

\textit{Methods}--High quality single crystals of \SRO were grown by the floating-zone method (see SM~\cite{supplement_2025} and~\cite{bobowski_improved_2019}).
The crystal was cut into several similarly sized disks with top and bottom surfaces parallel to the (100) plane.
The sample used in the main experiment was an uncleaved, oval-shaped crystal (3.5$\times$4 mm$^2$, 1.5 mm thick).
The $\mu$SR measurements were performed using the Flexible-Advanced-MuSR-Environment (FLAME) at the Paul Scherrer Institute (PSI), operational since 2022. 
The raw $\mu$SR data are available in Ref.~\cite{muSR_2025}. 
Key features enabling precise Knight shift measurements include: (i) high spatial homogeneity and temporal stability of the muon beam and magnetic field; (ii) the capability to switch between the measurement sample and a reference Ag plate without disturbing temperature or field conditions. 
Schematics of the FLAME setup, tandem sample holder with the \SRO crystal, and the Ag reference plate are shown in Figs.~\ref{fig2}(a-b).\\

\textit{Contributions to the muon Knight shift}--Here,
we outline the contributions to the muon Knight shift as illustrated in Fig.~\ref{fig1}; more details are given in Section S4 of SM~\cite{supplement_2025}.
In a $\mu$SR experiment, the Knight shift $K\subm{obs}$ represents the difference between the external magnetic field $B\subm{ext}$ and the local field $B\subm{obs}$ experienced by muons:   
\begin{equation}
K\subm{obs}=(B\subm{obs}/B\subm{ext})-1.
\label{eq1}\\
\end{equation}
\noindent Here, $B\subm{obs}$ is obtained from the precession frequency of an implanted positive muon, $\omega$=$\gamma B\subm{obs}$ with $\gamma/2\pi$=135.53 MHz/T, and reflects the effective local field at the muon site under $B\subm{ext}$. 
$K\subm{obs}$ consists of geometric and microscopic contributions. 
The geometric terms, arising from the demagnetizing and Lorentz fields \cite{amato_introduction_2024}, depend only on the sample shape and magnetization and are subtracted from $K\subm{obs}$.
The microscopic terms $K\subm{\mu}$, containing orbital and spin terms, reflect intrinsic material properties.
The orbital term $K\subm{orbital}$ includes
orbital dipolar terms originating from Meissner screening, Langevin core diamagnetism, and Van Vleck paramagnetism.
The spin term divides into dipolar $K\subm{spin\text{-}dipolar}$ and contact $K\subm{spin\text{-}contact}$ contributions: 
$K\subm{spin\text{-}dipolar}$ in materials without magnetic ions reflects the Pauli spin susceptibility in the normal state and the Cooper-pair spin susceptibility in the superconducting state, 
while $K\subm{spin\text{-}contact}$ arises from hyperfine interaction between the muon and electron spins.
They directly probe spin susceptibility and reflect superconducting pairing symmetry. 
Since the muon Knight shift signal (including $K\subm{spin\text{-}contact}$) is intrinsically small for $d$-electron-based superconductors compared with NMR (see Table~I in SM~\cite{supplement_2025}), extracting $K\subm{spin\text{-}contact}$ from $K\subm{\mu}$ is not straightforward.

To overcome this, we introduce a protocol combining the DC magnetic susceptibility $\chi=M/H$ measured with a superconducting-quantum-interference-device (SQUID) magnetometer, which also consists of geometric and dipolar contributions. 
The dipolar term is common to those for the muon Knight shift, $K\subm{dipolar} = K\subm{orbital\text{-}dipolar} + K\subm{spin\text{-}dipolar}$.
By properly subtracting the dipolar contribution from $K\subm{\mu}$, one can extract $K\subm{spin\text{-}contact}$, absent in the SQUID signal.
Thus, decomposition of dipolar contributions into orbital and spin parts is unnecessary to obtain the spin susceptibility in the normal and superconducting states.\\

\textit{Multiple disks vs single disk}--In many earlier $\mu$SR experiments, multiple crystal pieces were mounted adjacent to one another to cover the muon-beam's cross-section, typically about 1~cm in diameter.
However, we realized that such an arrangement is unsuitable for transverse-field measurements under low but finite magnetic fields. 
Stray fields from the Meissner effect in neighboring crystals add to the applied field and alter the local field sensed by muons, producing an apparent paramagnetic Knight shift in the superconducting state. 
This artifact also appears in the normal state but is reduced by a factor of $\chi\subm{normal}/\chi\subm{super}$.

Figures~\ref{fig2}(c-d) compare the Fourier transforms of the muon time spectra for the two sample configurations at 0.2~T. 
Figure~\ref{fig2}(e) shows the temperature-dependence of $K\subm{obs}$ in \SRO under an in-plane  magnetic field of 0.2~T along [100] direction for two configurations: six-crystals (red circles) and one-crystal (black circles). 
The red and black squares denote the Ag reference used for field calibration.
$T\subm{c}$ of the \SRO crystal is 1.48 K at zero-field (see SM~\cite{supplement_2025}).
Since the Knight shift of Ag, $K\subm{Ag}=+110\pm11$ ppm~\cite{hotz_experimental_2023}, serves as the field standard, $K\subm{obs}$ of \SRO is derived as:
\begin{equation} 
   1 + K\subm{obs} = ({B\subm{obs}}/{B\subm{Ag}})(1 + K\subm{Ag}).
   \label{eq2}
\end{equation}

\begin{figure*}[t]
    \begin{center}
    \includegraphics[scale=0.95]{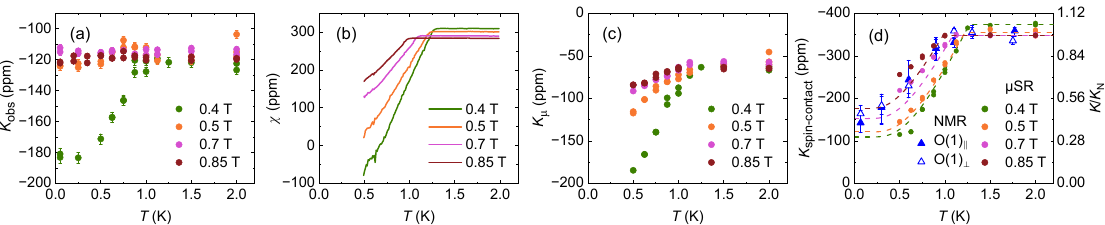}
    \caption{(a-b) Observed muon Knight shift $K\subm{obs}$ and DC magnetic susceptibility $\chi$ of the same \SRO crystal.
    (a) Temperature-dependence of $K\subm{obs}$ in $B\subm{ext}=0.4\text{-}0.85$~T along [100], calibrated using the Ag reference. 
    (b) Temperature-dependence of $\chi=M/H$ in $B\subm{ext}=0.4\text{-}0.85$~T along [100] 
    measured on warming from 0.5~K to 2~K after field-cooling from 3~K. 
    $B=\mu_0(M+H)$, where $M$ and $H$ are in A/m.
    (c-d) Derived intrinsic Knight shifts of \SRO under various magnetic fields. 
    (c) $K\subm{\mu}$ obtained from $K\subm{obs}$ by removing the contribution of Lorentz and demagnetizing field contributions.
    (d) Spin Knight shift represented by $K\subm{spin\text{-}contact}$. 
    The vertical scale is inverted in (d), reflecting the negative coupling between dipolar and contact fields of spin susceptibility.
    For comparison, the NMR Knight shift at 0.6916 T along [100] is also shown \cite{ishida_reduction_2020}. 
    O(1)$_\parallel$ and O(1)$_\perp$ represent two distinct planer O(1) sites. 
    Excellent agreement is observed between NMR and $\mu$SR results at 0.7 T.
    }
    \label{fig3}
    \end{center}
\end{figure*}
The measurement using one-crystal exhibits the expected diamagnetic response of the Knight shift below $T\subm{c}$, primarily reflecting the orbital diamagnetic shift ($K\subm{dipolar,Mei}$) associated with vortex-state screening.
In contrast, the measurement using six-crystals exhibits an anomalous increase in the Knight shift below $T\subm{c}$, a nonphysical result.
Notably, a stray-field as small as 0.1~mT in an applied field of 0.2~T corresponds to a substantial shift of 500~ppm in $K\subm{obs}$, as indicated by the right axis of Fig.~\ref{fig2}(e). For this reason, all subsequent Knight shift measurements reported are conducted using one-crystal. 
Details of the $\mu$SR spectra, including the asymmetry-time spectra, and exponential ($\Lambda$) and Gaussian ($\sigma$) relaxation rates that characterize the internal-field inhomogeneity, are discussed in SM~\cite{supplement_2025}. 
Such artifacts do not affect zero-field $\mu$SR measurements, including those probing spontaneous TRSB, or high-field measurements where Meissner screening becomes negligible.

We note that in Fig.~\ref{fig2}(c) the characteristic spectral shape reflecting the field distribution in a vortex lattice, commonly known as the Redfield pattern, was not observed. 
The peak width is about 1 mT even in the normal state, mainly due to nuclear contributions. 
This is much broader than the expected width of 20~$\mu$T \cite{amano_pauli_2015} and the very long penetration depth of about 10~$\mu$m along the $c$-axis. 
The Redfield patterns in \SRO have been reported for fields parallel to the $c$-axis by $\mathrm{\mu}$SR \cite{aegerter_evidence_1998, yakovlev_atypical_2025, khasanov_-plane_2023}, where the relevant penetration depth is only 0.2 $\mu$m.\\

\textit{Temperature-dependent Knight shift}--Figure~\ref{fig3}(a) summarizes the temperature-dependent Knight shift $K\subm{obs}$ of the \SRO crystal under different magnetic fields applied along [100].
Full details of $K\subm{obs}$ are presented in SM~\cite{supplement_2025}, where
$K\subm{obs}$ of \SRO in the normal state at 2.0~K is determined to be $-116\pm7$~ppm.
A strong reduction of $K\subm{obs}$ in \SRO observed below $T\subm{c}$ at 0.2 T (Fig. S3(a) in SM~\cite{supplement_2025}) is primarily attributed to the dipolar-Meissner diamagnetic response. 
With increasing magnetic field, the magnitude of this reduction becomes progressively smaller, as shown in Fig.~\ref{fig3}(a).\\

\textit{DC magnetization}--To extract $K_\mu$ and $K\subm{spin\text{-}contact}$ from $K\subm{obs}$, we independently determine $K\subm{dipolar}$ from temperature-dependent DC magnetization $M$, measured during warming from 0.5 to 2.0~K after field-cooling (FC).
These measurements were performed using the same \SRO crystal used in the $\mu$SR experiments, in a SQUID magnetometer equipped with an $^3$He refrigerator insert.
As shown in Fig.~\ref{fig1} and Eq.~S5 in SM~\cite{supplement_2025}, the difference between $\chi\subm{obs}$ and $\chi$ due to demagnetizing correction is negligible
in the SQUID measurement above 0.2~T.
Below 2~K, we consider that $K\subm{dipolar}=\chi$, where $\chi$ is the dimensionless DC magnetic susceptibility (see SM~\cite{supplement_2025}). 
In $K\subm{dipolar}$, the Pauli susceptibility of \SRO is nearly temperature-independent at low-temperature above $T\subm{c}$~\cite{maeno_two-dimensional_1997}, and other contributions
are expected to be substantially smaller \cite{udagawa_correlation_2010}.
Figure~\ref{fig3}(b) shows the temperature-dependence of $\chi$, exhibiting a clear diamagnetic response from orbital Meissner screening.

Reflecting the quasi-two-dimensional electronic structure of \SRO, $T\subm{c}$ in a magnetic field is highly sensitive to out-of-plane tilting.
To ensure accurate field alignment, we measure the onset temperature of superconductivity as a function of applied field and compare it with results by Kittaka $et~al$.~\cite{kittaka_angular_2009} as shown in SM~\cite{supplement_2025}. 
From this comparison, the misalignment angle relative to the $ab$-plane is estimated to be between 1 and 2 degrees in SQUID measurements, consistent with the alignment accuracy in our $\mu$SR measurements, as verified by the temperature-dependent $\Lambda$ (see SM~\cite{supplement_2025}). We note that $M(T)$ strongly depends on the measurement sequences: FC or zero-field-cooling (ZFC) (see SM~\cite{supplement_2025}), therefore it is crucial to perform magnetization measurements under FC conditions identical to the $\mu$SR measurements when extracting $K\subm{spin-contact}$ from $K_\mu$.
The vortex structure and magnetization of a Pauli-limited layered superconductor, used to model \SRO, have been theoretically calculated for fields parallel to the $ab$-plane, including tilt angles of 1 and 2 degrees~\cite{amano_pauli_2015}.
However, the theoretical magnetization overestimates the experimental FC values and yields nonphysical results if substituted directly for the experimental data. 
Therefore, we rely on experimentally determined FC magnetization to extract $K\subm{spin\text{-}contact}$.\\

\textit{Muon Knight shift in the superconducting state}--We analyze the muon Knight shift to extract $K\subm{spin\text{-}contact}$. The DC magnetic susceptibility $\chi=K\subm{dipolar}$ contains two primary contributions: a Pauli paramagnetic term dominating above $T\subm{c}$, and a diamagnetic response emerging below $T\subm{c}$. 
The paramagnetic component decreases below $T\subm{c}$ if the Cooper-pair spin susceptibility is reduced relative to the normal state susceptibility, as expected for spin-singlet-like pairing or for spin-triplet pairing with an external field applied parallel to the $d$-vector. 
This behavior is described by the Yosida function~\cite{mackenzie_superconductivity_2003,yosida_paramagnetic_1958}. 
By subtracting $K\subm{dipolar}$ measured by SQUID from $K\subm{\mu}$, we isolate $K\subm{spin\text{-}contact}$ arising from the contact interaction between the muon spin and polarized conduction electron spins.

Using the demagnetizing factor $N$ = 0.526 estimated for the sample shape used in both $\mu$SR and DC magnetization measurements~\cite{sato_simple_1989}, we calculate the Lorentz and demagnetizing field corrections as $B\subm{Lor}+B\subm{demag}=(1/3-N)M$. 
Applying this correction to the experimentally measured Knight shift $K\subm{obs}$ gives the intrinsic muon Knight shift $K\subm{\mu}$ from Eq.~\eqref{eq3}:
\begin{equation}
  K\subm{\mu} = K\subm{obs} + 0.193\chi. 
  \label{eq3}
\end{equation}
\noindent The resulting $K\subm{\mu}$ is shown in Fig.~\ref{fig3}(c). 
After removing $K\subm{dipolar}$ from $K_\mu$, we isolate $K\subm{spin\text{-}contact}$ under various magnetic fields, shown in Fig.~\ref{fig3}(d). 
We further compare these results with Yosida functions incorporating quasiparticle-spin contributions induced by magnetic fields, extracted from specific-heat measurements~\cite{deguchi_gap_2004, kittaka_searching_2018}, providing insight into residual spin polarization in \SRO.

The value of $K\subm{spin\text{-}contact}$ is estimated as $-357\pm11$~ppm in the normal state of \SRO.
Understanding the microscopic origin of this observed negative $K\subm{spin\text{-}contact}$ in the normal state remains a future issue of theoretical investigation.
At 0.2 T, this approach does not provide sufficient accuracy for $K\subm{spin\text{-}contact}$, as subtracting $\chi$ from $K_\mu$ leads to large uncertainty.
At higher fields, such as 0.4 T, the magnitude of the negative shift becomes sufficiently small. 
Moreover, a slight positive shift at 0.5 T below 1 K in $K\subm{obs}$ (Fig.~\ref{fig3}(a)) is attributable to a reduction of the spin term.
We note that $K\subm{obs}$ shown in Fig. \ref{fig3}(a) does not exhibit the negative shift expected for diamagnetic shifts of orbital and spin origin. 
This behavior arises because the contact field $K\subm{spin\text{-}contact}$ has the opposite sign to the dipolar field, causing near-cancellation above 0.5 T.\\

\textit{Conclusions}--We have demonstrated a solution to the long-standing challenge of determining the spin Knight shift of the $d$-electron-based unconventional superconductor \SRO using $\mu$SR.
Such an approach becomes possible through recent advancements in the $\mu$SR technique to accurately separate the sample spectrum from the reference signal used to monitor the external field.
We point out that sample arrangements often used in previous $\mu$SR measurements involving multiple samples may suffer from mutual stray fields in both normal and superconducting states.
By introducing a new protocol of combining DC magnetization measurements using the same sample under identical conditions, we successfully extracted $K\subm{spin\text{-}contact}$, representing the electron spin susceptibility. 

A significant reduction in the muon Knight shift below $T\subm{c}$ was observed over a broad range of magnetic fields, consistent with a spin-singlet-like decrease in spin susceptibility and extending the investigation beyond the NMR Knight shift at 0.7~T.
Since magnetization measurements were limited to 0.5~K, it is not possible in the present study to discriminate among fully gapped, line-nodal, helical, and inter-orbital superconducting states~\cite{suh_stabilizing_2020}.
However, these results motivate future experiments combining low-temperature SQUID and $\mu$SR measurements under high magnetic fields to explore the possible Fulde–Ferrell–Larkin–Ovchinnikov (FFLO) state~\cite{matsuda_fuldeferrelllarkinovchinnikov_2007}. With our newly achieved 7-ppm $\mu$SR Knight-shift resolution, it is now feasible to probe the FFLO state of Sr$_2$RuO$_4$, as reported by Kinjo $et~al.$~\cite{kinjo_superconducting_2022} using NMR, and similar states in other superconductors.
These results demonstrate the viability of $\mu$SR as a complementary technique to NMR for probing pairing symmetry in unconventional superconductors, particularly in highly conductive systems such as \SRO, which is essential to resolve ongoing debates about its pairing mechanism.\\

$Acknowledgements$--We would like to thank V. Grinenko,  W. Higemoto, M. Ichioka, K. Ishida, T. Johnson, S. Kitagawa, H-H. Klaus, H. Matsumura, Y. Okamoto, S. Paul, A. Ramires, and M. Sigirist for stimulating discussions, comments and useful information. We thank the PSI staff for their technical support during the muon experiment. 
Y.M. acknowledges support from the JSPS KAKENHI No. JP22H01168 and JP23K22439. 
H.M., A.I., and G.M. acknowledge support from the Kyoto University Foundation.
G.M. acknowledges support from the JSPS KAKENHI No. 25K17346, and Toyota Riken Scholar.
A.I. acknowledges support from JSPS KAKENHI No. JP24H01659.
Part of this work was performed using the facilities at the Institute for Solid State Physics, the University of Tokyo (ISSPkyodo-202501-GNBXX-0116).

\bibliography{muSR_PRL.bib}

\begin{thebibliography}{73}%
\makeatletter
\providecommand \@ifxundefined [1]{%
 \@ifx{#1\undefined}
}%
\providecommand \@ifnum [1]{%
 \ifnum #1\expandafter \@firstoftwo
 \else \expandafter \@secondoftwo
 \fi
}%
\providecommand \@ifx [1]{%
 \ifx #1\expandafter \@firstoftwo
 \else \expandafter \@secondoftwo
 \fi
}%
\providecommand \natexlab [1]{#1}%
\providecommand \enquote  [1]{``#1''}%
\providecommand \bibnamefont  [1]{#1}%
\providecommand \bibfnamefont [1]{#1}%
\providecommand \citenamefont [1]{#1}%
\providecommand \href@noop [0]{\@secondoftwo}%
\providecommand \href [0]{\begingroup \@sanitize@url \@href}%
\providecommand \@href[1]{\@@startlink{#1}\@@href}%
\providecommand \@@href[1]{\endgroup#1\@@endlink}%
\providecommand \@sanitize@url [0]{\catcode `\\12\catcode `\$12\catcode `\&12\catcode `\#12\catcode `\^12\catcode `\_12\catcode `\%12\relax}%
\providecommand \@@startlink[1]{}%
\providecommand \@@endlink[0]{}%
\providecommand \url  [0]{\begingroup\@sanitize@url \@url }%
\providecommand \@url [1]{\endgroup\@href {#1}{\urlprefix }}%
\providecommand \urlprefix  [0]{URL }%
\providecommand \Eprint [0]{\href }%
\providecommand \doibase [0]{https://doi.org/}%
\providecommand \selectlanguage [0]{\@gobble}%
\providecommand \bibinfo  [0]{\@secondoftwo}%
\providecommand \bibfield  [0]{\@secondoftwo}%
\providecommand \translation [1]{[#1]}%
\providecommand \BibitemOpen [0]{}%
\providecommand \bibitemStop [0]{}%
\providecommand \bibitemNoStop [0]{.\EOS\space}%
\providecommand \EOS [0]{\spacefactor3000\relax}%
\providecommand \BibitemShut  [1]{\csname bibitem#1\endcsname}%
\let\auto@bib@innerbib\@empty
%</preamble>
\bibitem [{\citenamefont {Maeno}\ \emph {et~al.}(1994)\citenamefont {Maeno}, \citenamefont {Hashimoto}, \citenamefont {Yoshida}, \citenamefont {Nishizaki}, \citenamefont {Fujita}, \citenamefont {Bednorz},\ and\ \citenamefont {Lichtenberg}}]{maeno_superconductivity_1994}%
  \BibitemOpen
  \bibfield  {author} {\bibinfo {author} {\bibfnamefont {Y.}~\bibnamefont {Maeno}}, \bibinfo {author} {\bibfnamefont {H.}~\bibnamefont {Hashimoto}}, \bibinfo {author} {\bibfnamefont {K.}~\bibnamefont {Yoshida}}, \bibinfo {author} {\bibfnamefont {S.}~\bibnamefont {Nishizaki}}, \bibinfo {author} {\bibfnamefont {T.}~\bibnamefont {Fujita}}, \bibinfo {author} {\bibfnamefont {J.~G.}\ \bibnamefont {Bednorz}},\ and\ \bibinfo {author} {\bibfnamefont {F.}~\bibnamefont {Lichtenberg}},\ }\href {https://doi.org/10.1038/372532a0} {\bibfield  {journal} {\bibinfo  {journal} {Nature}\ }\textbf {\bibinfo {volume} {372}},\ \bibinfo {pages} {532} (\bibinfo {year} {1994})}\BibitemShut {NoStop}%
\bibitem [{\citenamefont {Maeno}\ \emph {et~al.}(2024{\natexlab{a}})\citenamefont {Maeno}, \citenamefont {Ikeda},\ and\ \citenamefont {Mattoni}}]{maeno_thirty_2024}%
  \BibitemOpen
  \bibfield  {author} {\bibinfo {author} {\bibfnamefont {Y.}~\bibnamefont {Maeno}}, \bibinfo {author} {\bibfnamefont {A.}~\bibnamefont {Ikeda}},\ and\ \bibinfo {author} {\bibfnamefont {G.}~\bibnamefont {Mattoni}},\ }\href {https://doi.org/10.1038/s41567-024-02656-0} {\bibfield  {journal} {\bibinfo  {journal} {Nat. Phys.}\ }\textbf {\bibinfo {volume} {20}},\ \bibinfo {pages} {1712} (\bibinfo {year} {2024}{\natexlab{a}})}\BibitemShut {NoStop}%
\bibitem [{\citenamefont {Maeno}\ \emph {et~al.}(2024{\natexlab{b}})\citenamefont {Maeno}, \citenamefont {Yonezawa},\ and\ \citenamefont {Ramires}}]{maeno_still_2024}%
  \BibitemOpen
  \bibfield  {author} {\bibinfo {author} {\bibfnamefont {Y.}~\bibnamefont {Maeno}}, \bibinfo {author} {\bibfnamefont {S.}~\bibnamefont {Yonezawa}},\ and\ \bibinfo {author} {\bibfnamefont {A.}~\bibnamefont {Ramires}},\ }\href {https://doi.org/10.7566/JPSJ.93.062001} {\bibfield  {journal} {\bibinfo  {journal} {J. Phys. Soc. Jpn.}\ }\textbf {\bibinfo {volume} {93}},\ \bibinfo {pages} {062001} (\bibinfo {year} {2024}{\natexlab{b}})}\BibitemShut {NoStop}%
\bibitem [{\citenamefont {Knight}\ \emph {et~al.}(1956)\citenamefont {Knight}, \citenamefont {Androes},\ and\ \citenamefont {Hammond}}]{knight_nuclear_1956}%
  \BibitemOpen
  \bibfield  {author} {\bibinfo {author} {\bibfnamefont {W.~D.}\ \bibnamefont {Knight}}, \bibinfo {author} {\bibfnamefont {G.~M.}\ \bibnamefont {Androes}},\ and\ \bibinfo {author} {\bibfnamefont {R.~H.}\ \bibnamefont {Hammond}},\ }\href {https://doi.org/10.1103/PhysRev.104.852} {\bibfield  {journal} {\bibinfo  {journal} {Phys. Rev.}\ }\textbf {\bibinfo {volume} {104}},\ \bibinfo {pages} {852} (\bibinfo {year} {1956})}\BibitemShut {NoStop}%
\bibitem [{\citenamefont {Pustogow}\ \emph {et~al.}(2019)\citenamefont {Pustogow}, \citenamefont {Luo}, \citenamefont {Chronister}, \citenamefont {Su}, \citenamefont {Sokolov}, \citenamefont {Jerzembeck}, \citenamefont {Mackenzie}, \citenamefont {Hicks}, \citenamefont {Kikugawa}, \citenamefont {Raghu}, \citenamefont {Bauer},\ and\ \citenamefont {Brown}}]{pustogow_constraints_2019}%
  \BibitemOpen
  \bibfield  {author} {\bibinfo {author} {\bibfnamefont {A.}~\bibnamefont {Pustogow}}, \bibinfo {author} {\bibfnamefont {Y.}~\bibnamefont {Luo}}, \bibinfo {author} {\bibfnamefont {A.}~\bibnamefont {Chronister}}, \bibinfo {author} {\bibfnamefont {Y.-S.}\ \bibnamefont {Su}}, \bibinfo {author} {\bibfnamefont {D.~A.}\ \bibnamefont {Sokolov}}, \bibinfo {author} {\bibfnamefont {F.}~\bibnamefont {Jerzembeck}}, \bibinfo {author} {\bibfnamefont {A.~P.}\ \bibnamefont {Mackenzie}}, \bibinfo {author} {\bibfnamefont {C.~W.}\ \bibnamefont {Hicks}}, \bibinfo {author} {\bibfnamefont {N.}~\bibnamefont {Kikugawa}}, \bibinfo {author} {\bibfnamefont {S.}~\bibnamefont {Raghu}}, \bibinfo {author} {\bibfnamefont {E.~D.}\ \bibnamefont {Bauer}},\ and\ \bibinfo {author} {\bibfnamefont {S.~E.}\ \bibnamefont {Brown}},\ }\href {https://doi.org/10.1038/s41586-019-1596-2} {\bibfield  {journal} {\bibinfo  {journal} {Nature}\ }\textbf {\bibinfo {volume} {574}},\ \bibinfo {pages} {72} (\bibinfo {year} {2019})}\BibitemShut {NoStop}%
\bibitem [{\citenamefont {Ishida}\ \emph {et~al.}(2020)\citenamefont {Ishida}, \citenamefont {Manago}, \citenamefont {Kinjo},\ and\ \citenamefont {Maeno}}]{ishida_reduction_2020}%
  \BibitemOpen
  \bibfield  {author} {\bibinfo {author} {\bibfnamefont {K.}~\bibnamefont {Ishida}}, \bibinfo {author} {\bibfnamefont {M.}~\bibnamefont {Manago}}, \bibinfo {author} {\bibfnamefont {K.}~\bibnamefont {Kinjo}},\ and\ \bibinfo {author} {\bibfnamefont {Y.}~\bibnamefont {Maeno}},\ }\href {https://doi.org/10.7566/JPSJ.89.034712} {\bibfield  {journal} {\bibinfo  {journal} {J. Phys. Soc. Jpn.}\ }\textbf {\bibinfo {volume} {89}},\ \bibinfo {pages} {034712} (\bibinfo {year} {2020})}\BibitemShut {NoStop}%
\bibitem [{\citenamefont {Chronister}\ \emph {et~al.}(2021)\citenamefont {Chronister}, \citenamefont {Pustogow}, \citenamefont {Kikugawa}, \citenamefont {Sokolov}, \citenamefont {Jerzembeck}, \citenamefont {Hicks}, \citenamefont {Mackenzie}, \citenamefont {Bauer},\ and\ \citenamefont {Brown}}]{chronister_evidence_2021}%
  \BibitemOpen
  \bibfield  {author} {\bibinfo {author} {\bibfnamefont {A.}~\bibnamefont {Chronister}}, \bibinfo {author} {\bibfnamefont {A.}~\bibnamefont {Pustogow}}, \bibinfo {author} {\bibfnamefont {N.}~\bibnamefont {Kikugawa}}, \bibinfo {author} {\bibfnamefont {D.~A.}\ \bibnamefont {Sokolov}}, \bibinfo {author} {\bibfnamefont {F.}~\bibnamefont {Jerzembeck}}, \bibinfo {author} {\bibfnamefont {C.~W.}\ \bibnamefont {Hicks}}, \bibinfo {author} {\bibfnamefont {A.~P.}\ \bibnamefont {Mackenzie}}, \bibinfo {author} {\bibfnamefont {E.~D.}\ \bibnamefont {Bauer}},\ and\ \bibinfo {author} {\bibfnamefont {S.~E.}\ \bibnamefont {Brown}},\ }\href {https://doi.org/10.1073/pnas.2025313118} {\bibfield  {journal} {\bibinfo  {journal} {Proc. Natl. Acad. Sci. U.S.A.}\ }\textbf {\bibinfo {volume} {118}},\ \bibinfo {pages} {e2025313118} (\bibinfo {year} {2021})}\BibitemShut {NoStop}%
\bibitem [{\citenamefont {Petsch}\ \emph {et~al.}(2020)\citenamefont {Petsch}, \citenamefont {Zhu}, \citenamefont {Enderle}, \citenamefont {Mao}, \citenamefont {Maeno}, \citenamefont {Mazin},\ and\ \citenamefont {Hayden}}]{petsch_reduction_2020}%
  \BibitemOpen
  \bibfield  {author} {\bibinfo {author} {\bibfnamefont {A.}~\bibnamefont {Petsch}}, \bibinfo {author} {\bibfnamefont {M.}~\bibnamefont {Zhu}}, \bibinfo {author} {\bibfnamefont {M.}~\bibnamefont {Enderle}}, \bibinfo {author} {\bibfnamefont {Z.}~\bibnamefont {Mao}}, \bibinfo {author} {\bibfnamefont {Y.}~\bibnamefont {Maeno}}, \bibinfo {author} {\bibfnamefont {I.}~\bibnamefont {Mazin}},\ and\ \bibinfo {author} {\bibfnamefont {S.}~\bibnamefont {Hayden}},\ }\href {https://doi.org/10.1103/PhysRevLett.125.217004} {\bibfield  {journal} {\bibinfo  {journal} {Phys. Rev. Lett.}\ }\textbf {\bibinfo {volume} {125}},\ \bibinfo {pages} {217004} (\bibinfo {year} {2020})}\BibitemShut {NoStop}%
\bibitem [{\citenamefont {Amato}\ and\ \citenamefont {Morenzoni}(2024)}]{amato_introduction_2024}%
  \BibitemOpen
  \bibfield  {author} {\bibinfo {author} {\bibfnamefont {A.}~\bibnamefont {Amato}}\ and\ \bibinfo {author} {\bibfnamefont {E.}~\bibnamefont {Morenzoni}},\ }\href {https://doi.org/10.1007/978-3-031-44959-8} {\emph {\bibinfo {title} {Introduction to {Muon} {Spin} {Spectroscopy}}}},\ Vol.\ \bibinfo {volume} {961}\ (\bibinfo  {publisher} {Springer Cham},\ \bibinfo {year} {2024})\BibitemShut {NoStop}%
\bibitem [{\citenamefont {Higemoto}\ \emph {et~al.}(2016)\citenamefont {Higemoto}, \citenamefont {Aoki},\ and\ \citenamefont {MacLaughlin}}]{higemoto_spin_2016}%
  \BibitemOpen
  \bibfield  {author} {\bibinfo {author} {\bibfnamefont {W.}~\bibnamefont {Higemoto}}, \bibinfo {author} {\bibfnamefont {Y.}~\bibnamefont {Aoki}},\ and\ \bibinfo {author} {\bibfnamefont {D.~E.}\ \bibnamefont {MacLaughlin}},\ }\href {https://doi.org/10.7566/JPSJ.85.091007} {\bibfield  {journal} {\bibinfo  {journal} {J. Phys. Soc. Jpn.}\ }\textbf {\bibinfo {volume} {85}},\ \bibinfo {pages} {091007} (\bibinfo {year} {2016})}\BibitemShut {NoStop}%
\bibitem [{\citenamefont {Heffner}\ \emph {et~al.}(1986)\citenamefont {Heffner}, \citenamefont {Cooke}, \citenamefont {Fisk}, \citenamefont {Hutson}, \citenamefont {Schillaci}, \citenamefont {Smith}, \citenamefont {Willis}, \citenamefont {MacLaughlin}, \citenamefont {Boekema}, \citenamefont {Lichti}, \citenamefont {Denison},\ and\ \citenamefont {Oostens}}]{heffner_muon_1986}%
  \BibitemOpen
  \bibfield  {author} {\bibinfo {author} {\bibfnamefont {R.~H.}\ \bibnamefont {Heffner}}, \bibinfo {author} {\bibfnamefont {D.~W.}\ \bibnamefont {Cooke}}, \bibinfo {author} {\bibfnamefont {Z.}~\bibnamefont {Fisk}}, \bibinfo {author} {\bibfnamefont {R.~L.}\ \bibnamefont {Hutson}}, \bibinfo {author} {\bibfnamefont {M.~E.}\ \bibnamefont {Schillaci}}, \bibinfo {author} {\bibfnamefont {J.~L.}\ \bibnamefont {Smith}}, \bibinfo {author} {\bibfnamefont {J.~O.}\ \bibnamefont {Willis}}, \bibinfo {author} {\bibfnamefont {D.~E.}\ \bibnamefont {MacLaughlin}}, \bibinfo {author} {\bibfnamefont {C.}~\bibnamefont {Boekema}}, \bibinfo {author} {\bibfnamefont {R.~L.}\ \bibnamefont {Lichti}}, \bibinfo {author} {\bibfnamefont {A.~B.}\ \bibnamefont {Denison}},\ and\ \bibinfo {author} {\bibfnamefont {J.}~\bibnamefont {Oostens}},\ }\href {https://doi.org/10.1103/PhysRevLett.57.1255} {\bibfield  {journal} {\bibinfo  {journal} {Phys. Rev. Lett.}\ }\textbf {\bibinfo {volume} {57}},\ \bibinfo {pages} {1255} (\bibinfo {year}
  {1986})}\BibitemShut {NoStop}%
\bibitem [{\citenamefont {Uemura}\ \emph {et~al.}(1986)\citenamefont {Uemura}, \citenamefont {Kossler}, \citenamefont {Hitti}, \citenamefont {Kempton}, \citenamefont {Schone}, \citenamefont {Yu}, \citenamefont {Stronach}, \citenamefont {Lankford}, \citenamefont {Noakes}, \citenamefont {Keitel}, \citenamefont {Senba}, \citenamefont {Brewer}, \citenamefont {Ansaldo}, \citenamefont {Oonuki}, \citenamefont {Komatsubara}, \citenamefont {Aeppli}, \citenamefont {Bucher},\ and\ \citenamefont {Crow}}]{uemura_muon_1986}%
  \BibitemOpen
  \bibfield  {author} {\bibinfo {author} {\bibfnamefont {Y.~J.}\ \bibnamefont {Uemura}}, \bibinfo {author} {\bibfnamefont {W.~J.}\ \bibnamefont {Kossler}}, \bibinfo {author} {\bibfnamefont {B.}~\bibnamefont {Hitti}}, \bibinfo {author} {\bibfnamefont {J.~R.}\ \bibnamefont {Kempton}}, \bibinfo {author} {\bibfnamefont {H.~E.}\ \bibnamefont {Schone}}, \bibinfo {author} {\bibfnamefont {X.~H.}\ \bibnamefont {Yu}}, \bibinfo {author} {\bibfnamefont {C.~E.}\ \bibnamefont {Stronach}}, \bibinfo {author} {\bibfnamefont {W.~F.}\ \bibnamefont {Lankford}}, \bibinfo {author} {\bibfnamefont {D.~R.}\ \bibnamefont {Noakes}}, \bibinfo {author} {\bibfnamefont {R.}~\bibnamefont {Keitel}}, \bibinfo {author} {\bibfnamefont {M.}~\bibnamefont {Senba}}, \bibinfo {author} {\bibfnamefont {J.~H.}\ \bibnamefont {Brewer}}, \bibinfo {author} {\bibfnamefont {E.~J.}\ \bibnamefont {Ansaldo}}, \bibinfo {author} {\bibfnamefont {Y.}~\bibnamefont {Oonuki}}, \bibinfo {author} {\bibfnamefont {T.}~\bibnamefont {Komatsubara}}, \bibinfo {author}
  {\bibfnamefont {G.}~\bibnamefont {Aeppli}}, \bibinfo {author} {\bibfnamefont {E.}~\bibnamefont {Bucher}},\ and\ \bibinfo {author} {\bibfnamefont {J.~E.}\ \bibnamefont {Crow}},\ }\href {https://doi.org/10.1007/BF02401589} {\bibfield  {journal} {\bibinfo  {journal} {Hyperfine Interact.}\ }\textbf {\bibinfo {volume} {31}},\ \bibinfo {pages} {413} (\bibinfo {year} {1986})}\BibitemShut {NoStop}%
\bibitem [{\citenamefont {Luke}\ \emph {et~al.}(1991)\citenamefont {Luke}, \citenamefont {Le}, \citenamefont {Sternlieb}, \citenamefont {Wu}, \citenamefont {Uemura}, \citenamefont {Brewer}, \citenamefont {Kadono}, \citenamefont {Kiefl}, \citenamefont {Kreitzman}, \citenamefont {Riseman}, \citenamefont {Dalichaouch}, \citenamefont {Lee}, \citenamefont {Maple}, \citenamefont {Seaman}, \citenamefont {Armstrong}, \citenamefont {Ellis}, \citenamefont {Fisk},\ and\ \citenamefont {Smith}}]{luke_positive_1991}%
  \BibitemOpen
  \bibfield  {author} {\bibinfo {author} {\bibfnamefont {G.}~\bibnamefont {Luke}}, \bibinfo {author} {\bibfnamefont {L.}~\bibnamefont {Le}}, \bibinfo {author} {\bibfnamefont {B.}~\bibnamefont {Sternlieb}}, \bibinfo {author} {\bibfnamefont {W.}~\bibnamefont {Wu}}, \bibinfo {author} {\bibfnamefont {Y.}~\bibnamefont {Uemura}}, \bibinfo {author} {\bibfnamefont {J.}~\bibnamefont {Brewer}}, \bibinfo {author} {\bibfnamefont {R.}~\bibnamefont {Kadono}}, \bibinfo {author} {\bibfnamefont {R.}~\bibnamefont {Kiefl}}, \bibinfo {author} {\bibfnamefont {S.}~\bibnamefont {Kreitzman}}, \bibinfo {author} {\bibfnamefont {T.}~\bibnamefont {Riseman}}, \bibinfo {author} {\bibfnamefont {Y.}~\bibnamefont {Dalichaouch}}, \bibinfo {author} {\bibfnamefont {B.}~\bibnamefont {Lee}}, \bibinfo {author} {\bibfnamefont {M.}~\bibnamefont {Maple}}, \bibinfo {author} {\bibfnamefont {C.}~\bibnamefont {Seaman}}, \bibinfo {author} {\bibfnamefont {P.}~\bibnamefont {Armstrong}}, \bibinfo {author} {\bibfnamefont {R.}~\bibnamefont {Ellis}}, \bibinfo
  {author} {\bibfnamefont {Z.}~\bibnamefont {Fisk}},\ and\ \bibinfo {author} {\bibfnamefont {J.}~\bibnamefont {Smith}},\ }\href {https://doi.org/10.1016/0375-9601(91)90094-O} {\bibfield  {journal} {\bibinfo  {journal} {Phys. Lett. A}\ }\textbf {\bibinfo {volume} {157}},\ \bibinfo {pages} {173} (\bibinfo {year} {1991})}\BibitemShut {NoStop}%
\bibitem [{\citenamefont {Feyerherm}\ \emph {et~al.}(1994)\citenamefont {Feyerherm}, \citenamefont {Amato}, \citenamefont {Geibel}, \citenamefont {Gygax}, \citenamefont {Komatsubara}, \citenamefont {Sato}, \citenamefont {Schenck},\ and\ \citenamefont {Steglich}}]{feyerherm_sr_1994}%
  \BibitemOpen
  \bibfield  {author} {\bibinfo {author} {\bibfnamefont {R.}~\bibnamefont {Feyerherm}}, \bibinfo {author} {\bibfnamefont {A.}~\bibnamefont {Amato}}, \bibinfo {author} {\bibfnamefont {C.}~\bibnamefont {Geibel}}, \bibinfo {author} {\bibfnamefont {F.~N.}\ \bibnamefont {Gygax}}, \bibinfo {author} {\bibfnamefont {T.}~\bibnamefont {Komatsubara}}, \bibinfo {author} {\bibfnamefont {N.}~\bibnamefont {Sato}}, \bibinfo {author} {\bibfnamefont {A.}~\bibnamefont {Schenck}},\ and\ \bibinfo {author} {\bibfnamefont {F.}~\bibnamefont {Steglich}},\ }\href {https://doi.org/10.1016/0921-4526(94)91749-3} {\bibfield  {journal} {\bibinfo  {journal} {Physica B Condens. Matter}\ }\textbf {\bibinfo {volume} {199-200}},\ \bibinfo {pages} {103} (\bibinfo {year} {1994})}\BibitemShut {NoStop}%
\bibitem [{\citenamefont {Sonier}\ \emph {et~al.}(2000)\citenamefont {Sonier}, \citenamefont {Heffner}, \citenamefont {MacLaughlin}, \citenamefont {Smith}, \citenamefont {Cooley},\ and\ \citenamefont {Nieuwenhuys}}]{sonier_anomalous_2000}%
  \BibitemOpen
  \bibfield  {author} {\bibinfo {author} {\bibfnamefont {J.~E.}\ \bibnamefont {Sonier}}, \bibinfo {author} {\bibfnamefont {R.~H.}\ \bibnamefont {Heffner}}, \bibinfo {author} {\bibfnamefont {D.~E.}\ \bibnamefont {MacLaughlin}}, \bibinfo {author} {\bibfnamefont {J.~L.}\ \bibnamefont {Smith}}, \bibinfo {author} {\bibfnamefont {J.}~\bibnamefont {Cooley}},\ and\ \bibinfo {author} {\bibfnamefont {G.~J.}\ \bibnamefont {Nieuwenhuys}},\ }\href {https://doi.org/10.1016/S0921-4526(00)00234-9} {\bibfield  {journal} {\bibinfo  {journal} {Physica B Condens. Matter}\ }\textbf {\bibinfo {volume} {289-290}},\ \bibinfo {pages} {20} (\bibinfo {year} {2000})}\BibitemShut {NoStop}%
\bibitem [{\citenamefont {Koda}\ \emph {et~al.}(2002)\citenamefont {Koda}, \citenamefont {Higemoto}, \citenamefont {Kadono}, \citenamefont {Kawasaki}, \citenamefont {Ishida}, \citenamefont {Kitaoka}, \citenamefont {Geibel},\ and\ \citenamefont {Steglich}}]{koda_evidence_2002}%
  \BibitemOpen
  \bibfield  {author} {\bibinfo {author} {\bibfnamefont {A.}~\bibnamefont {Koda}}, \bibinfo {author} {\bibfnamefont {W.}~\bibnamefont {Higemoto}}, \bibinfo {author} {\bibfnamefont {R.}~\bibnamefont {Kadono}}, \bibinfo {author} {\bibfnamefont {Y.}~\bibnamefont {Kawasaki}}, \bibinfo {author} {\bibfnamefont {K.}~\bibnamefont {Ishida}}, \bibinfo {author} {\bibfnamefont {Y.}~\bibnamefont {Kitaoka}}, \bibinfo {author} {\bibfnamefont {C.}~\bibnamefont {Geibel}},\ and\ \bibinfo {author} {\bibfnamefont {F.}~\bibnamefont {Steglich}},\ }\href {https://doi.org/10.1143/JPSJ.71.1427} {\bibfield  {journal} {\bibinfo  {journal} {J. Phys. Soc. Jpn.}\ }\textbf {\bibinfo {volume} {71}},\ \bibinfo {pages} {1427} (\bibinfo {year} {2002})}\BibitemShut {NoStop}%
\bibitem [{\citenamefont {Sonier}\ \emph {et~al.}(2003)\citenamefont {Sonier}, \citenamefont {Poon}, \citenamefont {Luke}, \citenamefont {Kyriakou}, \citenamefont {Miller}, \citenamefont {Liang}, \citenamefont {Wiebe}, \citenamefont {Fournier},\ and\ \citenamefont {Greene}}]{sonier_superconductivity_2003}%
  \BibitemOpen
  \bibfield  {author} {\bibinfo {author} {\bibfnamefont {J.~E.}\ \bibnamefont {Sonier}}, \bibinfo {author} {\bibfnamefont {K.~F.}\ \bibnamefont {Poon}}, \bibinfo {author} {\bibfnamefont {G.~M.}\ \bibnamefont {Luke}}, \bibinfo {author} {\bibfnamefont {P.}~\bibnamefont {Kyriakou}}, \bibinfo {author} {\bibfnamefont {R.~I.}\ \bibnamefont {Miller}}, \bibinfo {author} {\bibfnamefont {R.}~\bibnamefont {Liang}}, \bibinfo {author} {\bibfnamefont {C.~R.}\ \bibnamefont {Wiebe}}, \bibinfo {author} {\bibfnamefont {P.}~\bibnamefont {Fournier}},\ and\ \bibinfo {author} {\bibfnamefont {R.~L.}\ \bibnamefont {Greene}},\ }\href {https://doi.org/10.1103/PhysRevLett.91.147002} {\bibfield  {journal} {\bibinfo  {journal} {Phys. Rev. Lett.}\ }\textbf {\bibinfo {volume} {91}},\ \bibinfo {pages} {147002} (\bibinfo {year} {2003})}\BibitemShut {NoStop}%
\bibitem [{\citenamefont {Higemoto}\ \emph {et~al.}(2006)\citenamefont {Higemoto}, \citenamefont {Haga}, \citenamefont {D.~Matsuda}, \citenamefont {Ōnuki}, \citenamefont {Ohishi}, \citenamefont {U.~Ito}, \citenamefont {Koda}, \citenamefont {R.~Saha},\ and\ \citenamefont {Kadono}}]{higemoto_possible_2006}%
  \BibitemOpen
  \bibfield  {author} {\bibinfo {author} {\bibfnamefont {W.}~\bibnamefont {Higemoto}}, \bibinfo {author} {\bibfnamefont {Y.}~\bibnamefont {Haga}}, \bibinfo {author} {\bibfnamefont {T.}~\bibnamefont {D.~Matsuda}}, \bibinfo {author} {\bibfnamefont {Y.}~\bibnamefont {Ōnuki}}, \bibinfo {author} {\bibfnamefont {K.}~\bibnamefont {Ohishi}}, \bibinfo {author} {\bibfnamefont {T.}~\bibnamefont {U.~Ito}}, \bibinfo {author} {\bibfnamefont {A.}~\bibnamefont {Koda}}, \bibinfo {author} {\bibfnamefont {S.}~\bibnamefont {R.~Saha}},\ and\ \bibinfo {author} {\bibfnamefont {R.}~\bibnamefont {Kadono}},\ }\href {https://doi.org/10.1143/JPSJ.75.124713} {\bibfield  {journal} {\bibinfo  {journal} {J. Phys. Soc. Jpn.}\ }\textbf {\bibinfo {volume} {75}},\ \bibinfo {pages} {124713} (\bibinfo {year} {2006})}\BibitemShut {NoStop}%
\bibitem [{\citenamefont {Higemoto}\ \emph {et~al.}(2007)\citenamefont {Higemoto}, \citenamefont {Saha}, \citenamefont {Koda}, \citenamefont {Ohishi}, \citenamefont {Kadono}, \citenamefont {Aoki}, \citenamefont {Sugawara},\ and\ \citenamefont {Sato}}]{higemoto_spin-triplet_2007}%
  \BibitemOpen
  \bibfield  {author} {\bibinfo {author} {\bibfnamefont {W.}~\bibnamefont {Higemoto}}, \bibinfo {author} {\bibfnamefont {S.~R.}\ \bibnamefont {Saha}}, \bibinfo {author} {\bibfnamefont {A.}~\bibnamefont {Koda}}, \bibinfo {author} {\bibfnamefont {K.}~\bibnamefont {Ohishi}}, \bibinfo {author} {\bibfnamefont {R.}~\bibnamefont {Kadono}}, \bibinfo {author} {\bibfnamefont {Y.}~\bibnamefont {Aoki}}, \bibinfo {author} {\bibfnamefont {H.}~\bibnamefont {Sugawara}},\ and\ \bibinfo {author} {\bibfnamefont {H.}~\bibnamefont {Sato}},\ }\href {https://doi.org/10.1103/PhysRevB.75.020510} {\bibfield  {journal} {\bibinfo  {journal} {Phys. Rev. B}\ }\textbf {\bibinfo {volume} {75}},\ \bibinfo {pages} {020510} (\bibinfo {year} {2007})}\BibitemShut {NoStop}%
\bibitem [{\citenamefont {Higemoto}\ \emph {et~al.}(2010)\citenamefont {Higemoto}, \citenamefont {Koda}, \citenamefont {Kadono}, \citenamefont {Ohishi}, \citenamefont {Haga}, \citenamefont {Shishido}, \citenamefont {Settai},\ and\ \citenamefont {Ōnuki}}]{higemoto_muon_2010}%
  \BibitemOpen
  \bibfield  {author} {\bibinfo {author} {\bibfnamefont {W.}~\bibnamefont {Higemoto}}, \bibinfo {author} {\bibfnamefont {A.}~\bibnamefont {Koda}}, \bibinfo {author} {\bibfnamefont {R.}~\bibnamefont {Kadono}}, \bibinfo {author} {\bibfnamefont {K.}~\bibnamefont {Ohishi}}, \bibinfo {author} {\bibfnamefont {Y.}~\bibnamefont {Haga}}, \bibinfo {author} {\bibfnamefont {H.}~\bibnamefont {Shishido}}, \bibinfo {author} {\bibfnamefont {R.}~\bibnamefont {Settai}},\ and\ \bibinfo {author} {\bibfnamefont {Y.}~\bibnamefont {Ōnuki}},\ }\href {https://doi.org/10.1088/1742-6596/225/1/012013} {\bibfield  {journal} {\bibinfo  {journal} {J. Phys.: Conf. Ser.}\ }\textbf {\bibinfo {volume} {225}},\ \bibinfo {pages} {012013} (\bibinfo {year} {2010})}\BibitemShut {NoStop}%
\bibitem [{\citenamefont {Azari}\ \emph {et~al.}(2023)\citenamefont {Azari}, \citenamefont {Goeks}, \citenamefont {Yakovlev}, \citenamefont {Abedi}, \citenamefont {Dunsiger}, \citenamefont {Thomas}, \citenamefont {Thompson}, \citenamefont {Rosa},\ and\ \citenamefont {Sonier}}]{azari__2023}%
  \BibitemOpen
  \bibfield  {author} {\bibinfo {author} {\bibfnamefont {N.}~\bibnamefont {Azari}}, \bibinfo {author} {\bibfnamefont {M.~R.}\ \bibnamefont {Goeks}}, \bibinfo {author} {\bibfnamefont {M.}~\bibnamefont {Yakovlev}}, \bibinfo {author} {\bibfnamefont {M.}~\bibnamefont {Abedi}}, \bibinfo {author} {\bibfnamefont {S.~R.}\ \bibnamefont {Dunsiger}}, \bibinfo {author} {\bibfnamefont {S.~M.}\ \bibnamefont {Thomas}}, \bibinfo {author} {\bibfnamefont {J.~D.}\ \bibnamefont {Thompson}}, \bibinfo {author} {\bibfnamefont {P.~F.~S.}\ \bibnamefont {Rosa}},\ and\ \bibinfo {author} {\bibfnamefont {J.~E.}\ \bibnamefont {Sonier}},\ }\href {https://doi.org/10.1103/PhysRevB.108.L081103} {\bibfield  {journal} {\bibinfo  {journal} {Phys. Rev. B}\ }\textbf {\bibinfo {volume} {108}},\ \bibinfo {pages} {L081103} (\bibinfo {year} {2023})}\BibitemShut {NoStop}%
\bibitem [{\citenamefont {Azari}\ \emph {et~al.}(2025)\citenamefont {Azari}, \citenamefont {Yakovlev}, \citenamefont {Dunsiger}, \citenamefont {Uzoh}, \citenamefont {Mun}, \citenamefont {Huddart}, \citenamefont {Blundell}, \citenamefont {Bordelon}, \citenamefont {Thomas}, \citenamefont {Thompson}, \citenamefont {Rosa},\ and\ \citenamefont {Sonier}}]{azari_coexistence_2025}%
  \BibitemOpen
  \bibfield  {author} {\bibinfo {author} {\bibfnamefont {N.}~\bibnamefont {Azari}}, \bibinfo {author} {\bibfnamefont {M.}~\bibnamefont {Yakovlev}}, \bibinfo {author} {\bibfnamefont {S.~R.}\ \bibnamefont {Dunsiger}}, \bibinfo {author} {\bibfnamefont {O.~P.}\ \bibnamefont {Uzoh}}, \bibinfo {author} {\bibfnamefont {E.}~\bibnamefont {Mun}}, \bibinfo {author} {\bibfnamefont {B.~M.}\ \bibnamefont {Huddart}}, \bibinfo {author} {\bibfnamefont {S.~J.}\ \bibnamefont {Blundell}}, \bibinfo {author} {\bibfnamefont {M.~M.}\ \bibnamefont {Bordelon}}, \bibinfo {author} {\bibfnamefont {S.~M.}\ \bibnamefont {Thomas}}, \bibinfo {author} {\bibfnamefont {J.~D.}\ \bibnamefont {Thompson}}, \bibinfo {author} {\bibfnamefont {P.~F.~S.}\ \bibnamefont {Rosa}},\ and\ \bibinfo {author} {\bibfnamefont {J.~E.}\ \bibnamefont {Sonier}},\ }\href {https://doi.org/10.1103/PhysRevB.111.014513} {\bibfield  {journal} {\bibinfo  {journal} {Phys. Rev. B}\ }\textbf {\bibinfo {volume} {111}},\ \bibinfo {pages} {014513} (\bibinfo {year}
  {2025})}\BibitemShut {NoStop}%
\bibitem [{\citenamefont {Das}\ \emph {et~al.}(2020)\citenamefont {Das}, \citenamefont {Kobayashi}, \citenamefont {Smylie}, \citenamefont {Mielke}, \citenamefont {Takahashi}, \citenamefont {Willa}, \citenamefont {Yin}, \citenamefont {Welp}, \citenamefont {Hasan}, \citenamefont {Amato}, \citenamefont {Luetkens},\ and\ \citenamefont {Guguchia}}]{das_time-reversal_2020}%
  \BibitemOpen
  \bibfield  {author} {\bibinfo {author} {\bibfnamefont {D.}~\bibnamefont {Das}}, \bibinfo {author} {\bibfnamefont {K.}~\bibnamefont {Kobayashi}}, \bibinfo {author} {\bibfnamefont {M.~P.}\ \bibnamefont {Smylie}}, \bibinfo {author} {\bibfnamefont {C.}~\bibnamefont {Mielke}}, \bibinfo {author} {\bibfnamefont {T.}~\bibnamefont {Takahashi}}, \bibinfo {author} {\bibfnamefont {K.}~\bibnamefont {Willa}}, \bibinfo {author} {\bibfnamefont {J.-X.}\ \bibnamefont {Yin}}, \bibinfo {author} {\bibfnamefont {U.}~\bibnamefont {Welp}}, \bibinfo {author} {\bibfnamefont {M.~Z.}\ \bibnamefont {Hasan}}, \bibinfo {author} {\bibfnamefont {A.}~\bibnamefont {Amato}}, \bibinfo {author} {\bibfnamefont {H.}~\bibnamefont {Luetkens}},\ and\ \bibinfo {author} {\bibfnamefont {Z.}~\bibnamefont {Guguchia}},\ }\href {https://doi.org/10.1103/PhysRevB.102.134514} {\bibfield  {journal} {\bibinfo  {journal} {Phys. Rev. B}\ }\textbf {\bibinfo {volume} {102}},\ \bibinfo {pages} {134514} (\bibinfo {year} {2020})}\BibitemShut {NoStop}%
\bibitem [{\citenamefont {Graham}\ \emph {et~al.}(2024)\citenamefont {Graham}, \citenamefont {Liu}, \citenamefont {Sazgari}, \citenamefont {Mielke~III}, \citenamefont {Medarde}, \citenamefont {Luetkens}, \citenamefont {Khasanov}, \citenamefont {Shi},\ and\ \citenamefont {Guguchia}}]{graham_microscopic_2024}%
  \BibitemOpen
  \bibfield  {author} {\bibinfo {author} {\bibfnamefont {J.~N.}\ \bibnamefont {Graham}}, \bibinfo {author} {\bibfnamefont {H.}~\bibnamefont {Liu}}, \bibinfo {author} {\bibfnamefont {V.}~\bibnamefont {Sazgari}}, \bibinfo {author} {\bibfnamefont {C.}~\bibnamefont {Mielke~III}}, \bibinfo {author} {\bibfnamefont {M.}~\bibnamefont {Medarde}}, \bibinfo {author} {\bibfnamefont {H.}~\bibnamefont {Luetkens}}, \bibinfo {author} {\bibfnamefont {R.}~\bibnamefont {Khasanov}}, \bibinfo {author} {\bibfnamefont {Y.}~\bibnamefont {Shi}},\ and\ \bibinfo {author} {\bibfnamefont {Z.}~\bibnamefont {Guguchia}},\ }\href {https://doi.org/10.1038/s43246-024-00666-2} {\bibfield  {journal} {\bibinfo  {journal} {Commun. Mater.}\ }\textbf {\bibinfo {volume} {5}},\ \bibinfo {pages} {1} (\bibinfo {year} {2024})}\BibitemShut {NoStop}%
\bibitem [{\citenamefont {Aegerter}\ \emph {et~al.}(1998)\citenamefont {Aegerter}, \citenamefont {Lloyd}, \citenamefont {Ager}, \citenamefont {Lee}, \citenamefont {Romer}, \citenamefont {Keller},\ and\ \citenamefont {Forgan}}]{aegerter_evidence_1998}%
  \BibitemOpen
  \bibfield  {author} {\bibinfo {author} {\bibfnamefont {C.~M.}\ \bibnamefont {Aegerter}}, \bibinfo {author} {\bibfnamefont {S.~H.}\ \bibnamefont {Lloyd}}, \bibinfo {author} {\bibfnamefont {C.}~\bibnamefont {Ager}}, \bibinfo {author} {\bibfnamefont {S.~L.}\ \bibnamefont {Lee}}, \bibinfo {author} {\bibfnamefont {S.}~\bibnamefont {Romer}}, \bibinfo {author} {\bibfnamefont {H.}~\bibnamefont {Keller}},\ and\ \bibinfo {author} {\bibfnamefont {E.~M.}\ \bibnamefont {Forgan}},\ }\href {https://doi.org/10.1088/0953-8984/10/33/013} {\bibfield  {journal} {\bibinfo  {journal} {J. Phys.: Condens. Matter}\ }\textbf {\bibinfo {volume} {10}},\ \bibinfo {pages} {7445} (\bibinfo {year} {1998})}\BibitemShut {NoStop}%
\bibitem [{\citenamefont {Maisuradze}\ \emph {et~al.}(2009)\citenamefont {Maisuradze}, \citenamefont {Khasanov}, \citenamefont {Shengelaya},\ and\ \citenamefont {Keller}}]{maisuradze_comparison_2009}%
  \BibitemOpen
  \bibfield  {author} {\bibinfo {author} {\bibfnamefont {A.}~\bibnamefont {Maisuradze}}, \bibinfo {author} {\bibfnamefont {R.}~\bibnamefont {Khasanov}}, \bibinfo {author} {\bibfnamefont {A.}~\bibnamefont {Shengelaya}},\ and\ \bibinfo {author} {\bibfnamefont {H.}~\bibnamefont {Keller}},\ }\href {https://doi.org/10.1088/0953-8984/21/7/075701} {\bibfield  {journal} {\bibinfo  {journal} {J. Phys.: Condens. Matter}\ }\textbf {\bibinfo {volume} {21}},\ \bibinfo {pages} {075701} (\bibinfo {year} {2009})}\BibitemShut {NoStop}%
\bibitem [{\citenamefont {Ray}\ \emph {et~al.}(2014)\citenamefont {Ray}, \citenamefont {Gibbs}, \citenamefont {Bending}, \citenamefont {Curran}, \citenamefont {Babaev}, \citenamefont {Baines}, \citenamefont {Mackenzie},\ and\ \citenamefont {Lee}}]{ray_muon-spin_2014}%
  \BibitemOpen
  \bibfield  {author} {\bibinfo {author} {\bibfnamefont {S.~J.}\ \bibnamefont {Ray}}, \bibinfo {author} {\bibfnamefont {A.~S.}\ \bibnamefont {Gibbs}}, \bibinfo {author} {\bibfnamefont {S.~J.}\ \bibnamefont {Bending}}, \bibinfo {author} {\bibfnamefont {P.~J.}\ \bibnamefont {Curran}}, \bibinfo {author} {\bibfnamefont {E.}~\bibnamefont {Babaev}}, \bibinfo {author} {\bibfnamefont {C.}~\bibnamefont {Baines}}, \bibinfo {author} {\bibfnamefont {A.~P.}\ \bibnamefont {Mackenzie}},\ and\ \bibinfo {author} {\bibfnamefont {S.~L.}\ \bibnamefont {Lee}},\ }\href {https://doi.org/10.1103/PhysRevB.89.094504} {\bibfield  {journal} {\bibinfo  {journal} {Phys. Rev. B}\ }\textbf {\bibinfo {volume} {89}},\ \bibinfo {pages} {094504} (\bibinfo {year} {2014})}\BibitemShut {NoStop}%
\bibitem [{\citenamefont {Yakovlev}\ \emph {et~al.}(2025)\citenamefont {Yakovlev}, \citenamefont {Kartsonas},\ and\ \citenamefont {Sonier}}]{yakovlev_atypical_2025}%
  \BibitemOpen
  \bibfield  {author} {\bibinfo {author} {\bibfnamefont {M.}~\bibnamefont {Yakovlev}}, \bibinfo {author} {\bibfnamefont {Z.}~\bibnamefont {Kartsonas}},\ and\ \bibinfo {author} {\bibfnamefont {J.~E.}\ \bibnamefont {Sonier}},\ }\href {https://doi.org/10.1103/PhysRevB.111.094509} {\bibfield  {journal} {\bibinfo  {journal} {Phys. Rev. B}\ }\textbf {\bibinfo {volume} {111}},\ \bibinfo {pages} {094509} (\bibinfo {year} {2025})}\BibitemShut {NoStop}%
\bibitem [{\citenamefont {Ikeda}\ \emph {et~al.}(2020)\citenamefont {Ikeda}, \citenamefont {Guguchia}, \citenamefont {Oudah}, \citenamefont {Koibuchi}, \citenamefont {Yonezawa}, \citenamefont {Das}, \citenamefont {Shiroka}, \citenamefont {Luetkens},\ and\ \citenamefont {Maeno}}]{ikeda_penetration_2020}%
  \BibitemOpen
  \bibfield  {author} {\bibinfo {author} {\bibfnamefont {A.}~\bibnamefont {Ikeda}}, \bibinfo {author} {\bibfnamefont {Z.}~\bibnamefont {Guguchia}}, \bibinfo {author} {\bibfnamefont {M.}~\bibnamefont {Oudah}}, \bibinfo {author} {\bibfnamefont {S.}~\bibnamefont {Koibuchi}}, \bibinfo {author} {\bibfnamefont {S.}~\bibnamefont {Yonezawa}}, \bibinfo {author} {\bibfnamefont {D.}~\bibnamefont {Das}}, \bibinfo {author} {\bibfnamefont {T.}~\bibnamefont {Shiroka}}, \bibinfo {author} {\bibfnamefont {H.}~\bibnamefont {Luetkens}},\ and\ \bibinfo {author} {\bibfnamefont {Y.}~\bibnamefont {Maeno}},\ }\href {https://doi.org/10.1103/PhysRevB.101.174503} {\bibfield  {journal} {\bibinfo  {journal} {Phys. Rev. B}\ }\textbf {\bibinfo {volume} {101}},\ \bibinfo {pages} {174503} (\bibinfo {year} {2020})}\BibitemShut {NoStop}%
\bibitem [{\citenamefont {Khasanov}\ \emph {et~al.}(2023)\citenamefont {Khasanov}, \citenamefont {Ramires}, \citenamefont {Grinenko}, \citenamefont {Shipulin}, \citenamefont {Kikugawa}, \citenamefont {Sokolov}, \citenamefont {Krieger}, \citenamefont {Hicken}, \citenamefont {Maeno}, \citenamefont {Luetkens},\ and\ \citenamefont {Guguchia}}]{khasanov_-plane_2023}%
  \BibitemOpen
  \bibfield  {author} {\bibinfo {author} {\bibfnamefont {R.}~\bibnamefont {Khasanov}}, \bibinfo {author} {\bibfnamefont {A.}~\bibnamefont {Ramires}}, \bibinfo {author} {\bibfnamefont {V.}~\bibnamefont {Grinenko}}, \bibinfo {author} {\bibfnamefont {I.}~\bibnamefont {Shipulin}}, \bibinfo {author} {\bibfnamefont {N.}~\bibnamefont {Kikugawa}}, \bibinfo {author} {\bibfnamefont {D.~A.}\ \bibnamefont {Sokolov}}, \bibinfo {author} {\bibfnamefont {J.~A.}\ \bibnamefont {Krieger}}, \bibinfo {author} {\bibfnamefont {T.~J.}\ \bibnamefont {Hicken}}, \bibinfo {author} {\bibfnamefont {Y.}~\bibnamefont {Maeno}}, \bibinfo {author} {\bibfnamefont {H.}~\bibnamefont {Luetkens}},\ and\ \bibinfo {author} {\bibfnamefont {Z.}~\bibnamefont {Guguchia}},\ }\href {https://doi.org/10.1103/PhysRevLett.131.236001} {\bibfield  {journal} {\bibinfo  {journal} {Phys. Rev. Lett.}\ }\textbf {\bibinfo {volume} {131}},\ \bibinfo {pages} {236001} (\bibinfo {year} {2023})}\BibitemShut {NoStop}%
\bibitem [{\citenamefont {Luke}\ \emph {et~al.}(1998)\citenamefont {Luke}, \citenamefont {Fudamoto}, \citenamefont {Kojima}, \citenamefont {Larkin}, \citenamefont {Merrin}, \citenamefont {Nachumi}, \citenamefont {Uemura}, \citenamefont {Maeno}, \citenamefont {Mao}, \citenamefont {Mori}, \citenamefont {Nakamura},\ and\ \citenamefont {Sigrist}}]{luke_time-reversal_1998}%
  \BibitemOpen
  \bibfield  {author} {\bibinfo {author} {\bibfnamefont {G.~M.}\ \bibnamefont {Luke}}, \bibinfo {author} {\bibfnamefont {Y.}~\bibnamefont {Fudamoto}}, \bibinfo {author} {\bibfnamefont {K.~M.}\ \bibnamefont {Kojima}}, \bibinfo {author} {\bibfnamefont {M.~I.}\ \bibnamefont {Larkin}}, \bibinfo {author} {\bibfnamefont {J.}~\bibnamefont {Merrin}}, \bibinfo {author} {\bibfnamefont {B.}~\bibnamefont {Nachumi}}, \bibinfo {author} {\bibfnamefont {Y.~J.}\ \bibnamefont {Uemura}}, \bibinfo {author} {\bibfnamefont {Y.}~\bibnamefont {Maeno}}, \bibinfo {author} {\bibfnamefont {Z.~Q.}\ \bibnamefont {Mao}}, \bibinfo {author} {\bibfnamefont {Y.}~\bibnamefont {Mori}}, \bibinfo {author} {\bibfnamefont {H.}~\bibnamefont {Nakamura}},\ and\ \bibinfo {author} {\bibfnamefont {M.}~\bibnamefont {Sigrist}},\ }\href {https://doi.org/10.1038/29038} {\bibfield  {journal} {\bibinfo  {journal} {Nature}\ }\textbf {\bibinfo {volume} {394}},\ \bibinfo {pages} {558} (\bibinfo {year} {1998})}\BibitemShut {NoStop}%
\bibitem [{\citenamefont {Luke}\ \emph {et~al.}(2000)\citenamefont {Luke}, \citenamefont {Fudamoto}, \citenamefont {Kojima}, \citenamefont {Larkin}, \citenamefont {Nachumi}, \citenamefont {Uemura}, \citenamefont {Sonier}, \citenamefont {Maeno}, \citenamefont {Mao}, \citenamefont {Mori},\ and\ \citenamefont {Agterberg}}]{luke_unconventional_2000}%
  \BibitemOpen
  \bibfield  {author} {\bibinfo {author} {\bibfnamefont {G.~M.}\ \bibnamefont {Luke}}, \bibinfo {author} {\bibfnamefont {Y.}~\bibnamefont {Fudamoto}}, \bibinfo {author} {\bibfnamefont {K.~M.}\ \bibnamefont {Kojima}}, \bibinfo {author} {\bibfnamefont {M.~I.}\ \bibnamefont {Larkin}}, \bibinfo {author} {\bibfnamefont {B.}~\bibnamefont {Nachumi}}, \bibinfo {author} {\bibfnamefont {Y.~J.}\ \bibnamefont {Uemura}}, \bibinfo {author} {\bibfnamefont {J.~E.}\ \bibnamefont {Sonier}}, \bibinfo {author} {\bibfnamefont {Y.}~\bibnamefont {Maeno}}, \bibinfo {author} {\bibfnamefont {Z.~Q.}\ \bibnamefont {Mao}}, \bibinfo {author} {\bibfnamefont {Y.}~\bibnamefont {Mori}},\ and\ \bibinfo {author} {\bibfnamefont {D.~F.}\ \bibnamefont {Agterberg}},\ }\href {https://doi.org/10.1016/S0921-4526(00)00414-2} {\bibfield  {journal} {\bibinfo  {journal} {Physica B: Condens. Matter}\ }\textbf {\bibinfo {volume} {289-290}},\ \bibinfo {pages} {373} (\bibinfo {year} {2000})}\BibitemShut {NoStop}%
\bibitem [{\citenamefont {Shiroka}\ \emph {et~al.}(2012)\citenamefont {Shiroka}, \citenamefont {Fittipaldi}, \citenamefont {Cuoco}, \citenamefont {De~Renzi}, \citenamefont {Maeno}, \citenamefont {Lycett}, \citenamefont {Ramos}, \citenamefont {Forgan}, \citenamefont {Baines}, \citenamefont {Rost}, \citenamefont {Granata},\ and\ \citenamefont {Vecchione}}]{shiroka_sr_2012}%
  \BibitemOpen
  \bibfield  {author} {\bibinfo {author} {\bibfnamefont {T.}~\bibnamefont {Shiroka}}, \bibinfo {author} {\bibfnamefont {R.}~\bibnamefont {Fittipaldi}}, \bibinfo {author} {\bibfnamefont {M.}~\bibnamefont {Cuoco}}, \bibinfo {author} {\bibfnamefont {R.}~\bibnamefont {De~Renzi}}, \bibinfo {author} {\bibfnamefont {Y.}~\bibnamefont {Maeno}}, \bibinfo {author} {\bibfnamefont {R.~J.}\ \bibnamefont {Lycett}}, \bibinfo {author} {\bibfnamefont {S.}~\bibnamefont {Ramos}}, \bibinfo {author} {\bibfnamefont {E.~M.}\ \bibnamefont {Forgan}}, \bibinfo {author} {\bibfnamefont {C.}~\bibnamefont {Baines}}, \bibinfo {author} {\bibfnamefont {A.}~\bibnamefont {Rost}}, \bibinfo {author} {\bibfnamefont {V.}~\bibnamefont {Granata}},\ and\ \bibinfo {author} {\bibfnamefont {A.}~\bibnamefont {Vecchione}},\ }\href {https://doi.org/10.1103/PhysRevB.85.134527} {\bibfield  {journal} {\bibinfo  {journal} {Phys. Rev. B}\ }\textbf {\bibinfo {volume} {85}},\ \bibinfo {pages} {134527} (\bibinfo {year} {2012})}\BibitemShut {NoStop}%
\bibitem [{\citenamefont {Grinenko}\ \emph {et~al.}(2021{\natexlab{a}})\citenamefont {Grinenko}, \citenamefont {Das}, \citenamefont {Gupta}, \citenamefont {Zinkl}, \citenamefont {Kikugawa}, \citenamefont {Maeno}, \citenamefont {Hicks}, \citenamefont {Klauss}, \citenamefont {Sigrist},\ and\ \citenamefont {Khasanov}}]{grinenko_unsplit_2021}%
  \BibitemOpen
  \bibfield  {author} {\bibinfo {author} {\bibfnamefont {V.}~\bibnamefont {Grinenko}}, \bibinfo {author} {\bibfnamefont {D.}~\bibnamefont {Das}}, \bibinfo {author} {\bibfnamefont {R.}~\bibnamefont {Gupta}}, \bibinfo {author} {\bibfnamefont {B.}~\bibnamefont {Zinkl}}, \bibinfo {author} {\bibfnamefont {N.}~\bibnamefont {Kikugawa}}, \bibinfo {author} {\bibfnamefont {Y.}~\bibnamefont {Maeno}}, \bibinfo {author} {\bibfnamefont {C.~W.}\ \bibnamefont {Hicks}}, \bibinfo {author} {\bibfnamefont {H.-H.}\ \bibnamefont {Klauss}}, \bibinfo {author} {\bibfnamefont {M.}~\bibnamefont {Sigrist}},\ and\ \bibinfo {author} {\bibfnamefont {R.}~\bibnamefont {Khasanov}},\ }\href {https://doi.org/10.1038/s41467-021-24176-8} {\bibfield  {journal} {\bibinfo  {journal} {Nat. Commun.}\ }\textbf {\bibinfo {volume} {12}},\ \bibinfo {pages} {3920} (\bibinfo {year} {2021}{\natexlab{a}})}\BibitemShut {NoStop}%
\bibitem [{\citenamefont {Grinenko}\ \emph {et~al.}(2021{\natexlab{b}})\citenamefont {Grinenko}, \citenamefont {Ghosh}, \citenamefont {Sarkar}, \citenamefont {Orain}, \citenamefont {Nikitin}, \citenamefont {Elender}, \citenamefont {Das}, \citenamefont {Guguchia}, \citenamefont {Brückner}, \citenamefont {Barber}, \citenamefont {Park}, \citenamefont {Kikugawa}, \citenamefont {Sokolov}, \citenamefont {Bobowski}, \citenamefont {Miyoshi}, \citenamefont {Maeno}, \citenamefont {Mackenzie}, \citenamefont {Luetkens}, \citenamefont {Hicks},\ and\ \citenamefont {Klauss}}]{grinenko_split_2021}%
  \BibitemOpen
  \bibfield  {author} {\bibinfo {author} {\bibfnamefont {V.}~\bibnamefont {Grinenko}}, \bibinfo {author} {\bibfnamefont {S.}~\bibnamefont {Ghosh}}, \bibinfo {author} {\bibfnamefont {R.}~\bibnamefont {Sarkar}}, \bibinfo {author} {\bibfnamefont {J.-C.}\ \bibnamefont {Orain}}, \bibinfo {author} {\bibfnamefont {A.}~\bibnamefont {Nikitin}}, \bibinfo {author} {\bibfnamefont {M.}~\bibnamefont {Elender}}, \bibinfo {author} {\bibfnamefont {D.}~\bibnamefont {Das}}, \bibinfo {author} {\bibfnamefont {Z.}~\bibnamefont {Guguchia}}, \bibinfo {author} {\bibfnamefont {F.}~\bibnamefont {Brückner}}, \bibinfo {author} {\bibfnamefont {M.~E.}\ \bibnamefont {Barber}}, \bibinfo {author} {\bibfnamefont {J.}~\bibnamefont {Park}}, \bibinfo {author} {\bibfnamefont {N.}~\bibnamefont {Kikugawa}}, \bibinfo {author} {\bibfnamefont {D.~A.}\ \bibnamefont {Sokolov}}, \bibinfo {author} {\bibfnamefont {J.~S.}\ \bibnamefont {Bobowski}}, \bibinfo {author} {\bibfnamefont {T.}~\bibnamefont {Miyoshi}}, \bibinfo {author} {\bibfnamefont
  {Y.}~\bibnamefont {Maeno}}, \bibinfo {author} {\bibfnamefont {A.~P.}\ \bibnamefont {Mackenzie}}, \bibinfo {author} {\bibfnamefont {H.}~\bibnamefont {Luetkens}}, \bibinfo {author} {\bibfnamefont {C.~W.}\ \bibnamefont {Hicks}},\ and\ \bibinfo {author} {\bibfnamefont {H.-H.}\ \bibnamefont {Klauss}},\ }\href {https://doi.org/10.1038/s41567-021-01182-7} {\bibfield  {journal} {\bibinfo  {journal} {Nat. Phys.}\ }\textbf {\bibinfo {volume} {17}},\ \bibinfo {pages} {748} (\bibinfo {year} {2021}{\natexlab{b}})}\BibitemShut {NoStop}%
\bibitem [{\citenamefont {Grinenko}\ \emph {et~al.}(2023)\citenamefont {Grinenko}, \citenamefont {Sarkar}, \citenamefont {Ghosh}, \citenamefont {Das}, \citenamefont {Guguchia}, \citenamefont {Luetkens}, \citenamefont {Shipulin}, \citenamefont {Ramires}, \citenamefont {Kikugawa}, \citenamefont {Maeno}, \citenamefont {Ishida}, \citenamefont {Hicks},\ and\ \citenamefont {Klauss}}]{grinenko_sr_2023}%
  \BibitemOpen
  \bibfield  {author} {\bibinfo {author} {\bibfnamefont {V.}~\bibnamefont {Grinenko}}, \bibinfo {author} {\bibfnamefont {R.}~\bibnamefont {Sarkar}}, \bibinfo {author} {\bibfnamefont {S.}~\bibnamefont {Ghosh}}, \bibinfo {author} {\bibfnamefont {D.}~\bibnamefont {Das}}, \bibinfo {author} {\bibfnamefont {Z.}~\bibnamefont {Guguchia}}, \bibinfo {author} {\bibfnamefont {H.}~\bibnamefont {Luetkens}}, \bibinfo {author} {\bibfnamefont {I.}~\bibnamefont {Shipulin}}, \bibinfo {author} {\bibfnamefont {A.}~\bibnamefont {Ramires}}, \bibinfo {author} {\bibfnamefont {N.}~\bibnamefont {Kikugawa}}, \bibinfo {author} {\bibfnamefont {Y.}~\bibnamefont {Maeno}}, \bibinfo {author} {\bibfnamefont {K.}~\bibnamefont {Ishida}}, \bibinfo {author} {\bibfnamefont {C.~W.}\ \bibnamefont {Hicks}},\ and\ \bibinfo {author} {\bibfnamefont {H.-H.}\ \bibnamefont {Klauss}},\ }\href {https://doi.org/10.1103/PhysRevB.107.024508} {\bibfield  {journal} {\bibinfo  {journal} {Phys. Rev. B}\ }\textbf {\bibinfo {volume} {107}},\ \bibinfo {pages} {024508}
  (\bibinfo {year} {2023})}\BibitemShut {NoStop}%
\bibitem [{\citenamefont {Oudah}\ \emph {et~al.}(2024)\citenamefont {Oudah}, \citenamefont {Cai}, \citenamefont {Sanchez}, \citenamefont {Bannies}, \citenamefont {Aronson}, \citenamefont {Kojima},\ and\ \citenamefont {Bonn}}]{oudah_time-reversal_2024}%
  \BibitemOpen
  \bibfield  {author} {\bibinfo {author} {\bibfnamefont {M.}~\bibnamefont {Oudah}}, \bibinfo {author} {\bibfnamefont {Y.}~\bibnamefont {Cai}}, \bibinfo {author} {\bibfnamefont {M.~V. D.~T.}\ \bibnamefont {Sanchez}}, \bibinfo {author} {\bibfnamefont {J.}~\bibnamefont {Bannies}}, \bibinfo {author} {\bibfnamefont {M.~C.}\ \bibnamefont {Aronson}}, \bibinfo {author} {\bibfnamefont {K.~M.}\ \bibnamefont {Kojima}},\ and\ \bibinfo {author} {\bibfnamefont {D.~A.}\ \bibnamefont {Bonn}},\ }\href {https://doi.org/10.1103/PhysRevB.110.134524} {\bibfield  {journal} {\bibinfo  {journal} {Phys. Rev. B}\ }\textbf {\bibinfo {volume} {110}},\ \bibinfo {pages} {134524} (\bibinfo {year} {2024})}\BibitemShut {NoStop}%
\bibitem [{\citenamefont {Kiefl}\ \emph {et~al.}(1993)\citenamefont {Kiefl}, \citenamefont {MacFarlane}, \citenamefont {Chow}, \citenamefont {Dunsiger}, \citenamefont {Duty}, \citenamefont {Johnston}, \citenamefont {Schneider}, \citenamefont {Sonier}, \citenamefont {Brard}, \citenamefont {Strongin}, \citenamefont {Fischer},\ and\ \citenamefont {Smith}}]{kiefl_coherence_1993}%
  \BibitemOpen
  \bibfield  {author} {\bibinfo {author} {\bibfnamefont {R.~F.}\ \bibnamefont {Kiefl}}, \bibinfo {author} {\bibfnamefont {W.~A.}\ \bibnamefont {MacFarlane}}, \bibinfo {author} {\bibfnamefont {K.~H.}\ \bibnamefont {Chow}}, \bibinfo {author} {\bibfnamefont {S.}~\bibnamefont {Dunsiger}}, \bibinfo {author} {\bibfnamefont {T.~L.}\ \bibnamefont {Duty}}, \bibinfo {author} {\bibfnamefont {T.~M.~S.}\ \bibnamefont {Johnston}}, \bibinfo {author} {\bibfnamefont {J.~W.}\ \bibnamefont {Schneider}}, \bibinfo {author} {\bibfnamefont {J.}~\bibnamefont {Sonier}}, \bibinfo {author} {\bibfnamefont {L.}~\bibnamefont {Brard}}, \bibinfo {author} {\bibfnamefont {R.~M.}\ \bibnamefont {Strongin}}, \bibinfo {author} {\bibfnamefont {J.~E.}\ \bibnamefont {Fischer}},\ and\ \bibinfo {author} {\bibfnamefont {A.~B.}\ \bibnamefont {Smith}},\ }\href {https://doi.org/10.1103/PhysRevLett.70.3987} {\bibfield  {journal} {\bibinfo  {journal} {Phys. Rev. Lett.}\ }\textbf {\bibinfo {volume} {70}},\ \bibinfo {pages} {3987} (\bibinfo {year}
  {1993})}\BibitemShut {NoStop}%
\bibitem [{\citenamefont {Macfarlane}\ \emph {et~al.}(1994)\citenamefont {Macfarlane}, \citenamefont {Kiefl}, \citenamefont {Chow}, \citenamefont {Dunsiger}, \citenamefont {Duty}, \citenamefont {Johnston}, \citenamefont {Schneider}, \citenamefont {Sonier}, \citenamefont {Brard}, \citenamefont {Strongin}, \citenamefont {Fischer},\ and\ \citenamefont {Smith}}]{macfarlane_hebel-slichter_1994}%
  \BibitemOpen
  \bibfield  {author} {\bibinfo {author} {\bibfnamefont {W.~A.}\ \bibnamefont {Macfarlane}}, \bibinfo {author} {\bibfnamefont {R.~F.}\ \bibnamefont {Kiefl}}, \bibinfo {author} {\bibfnamefont {K.~H.}\ \bibnamefont {Chow}}, \bibinfo {author} {\bibfnamefont {S.}~\bibnamefont {Dunsiger}}, \bibinfo {author} {\bibfnamefont {T.~L.}\ \bibnamefont {Duty}}, \bibinfo {author} {\bibfnamefont {T.~M.~S.}\ \bibnamefont {Johnston}}, \bibinfo {author} {\bibfnamefont {J.~W.}\ \bibnamefont {Schneider}}, \bibinfo {author} {\bibfnamefont {J.}~\bibnamefont {Sonier}}, \bibinfo {author} {\bibfnamefont {L.}~\bibnamefont {Brard}}, \bibinfo {author} {\bibfnamefont {R.~M.}\ \bibnamefont {Strongin}}, \bibinfo {author} {\bibfnamefont {J.~E.}\ \bibnamefont {Fischer}},\ and\ \bibinfo {author} {\bibfnamefont {A.~B.}\ \bibnamefont {Smith}},\ }\href {https://doi.org/10.1007/BF02068935} {\bibfield  {journal} {\bibinfo  {journal} {Hyperfine Interact.}\ }\textbf {\bibinfo {volume} {86}},\ \bibinfo {pages} {467} (\bibinfo {year}
  {1994})}\BibitemShut {NoStop}%
\bibitem [{\citenamefont {Nishida}\ \emph {et~al.}(1987)\citenamefont {Nishida}, \citenamefont {Miyatake}, \citenamefont {Shimada}, \citenamefont {Okuma}, \citenamefont {Ishikawa}, \citenamefont {Takabatake}, \citenamefont {Nakazawa}, \citenamefont {Kuno}, \citenamefont {Keitel}, \citenamefont {Brewer}, \citenamefont {Riseman}, \citenamefont {Williams}, \citenamefont {Watanabe}, \citenamefont {Yamazaki}, \citenamefont {Nishiyama}, \citenamefont {Nagamine}, \citenamefont {Ansaldo},\ and\ \citenamefont {Torikai}}]{nishida_first_1987}%
  \BibitemOpen
  \bibfield  {author} {\bibinfo {author} {\bibfnamefont {N.}~\bibnamefont {Nishida}}, \bibinfo {author} {\bibfnamefont {H.}~\bibnamefont {Miyatake}}, \bibinfo {author} {\bibfnamefont {D.}~\bibnamefont {Shimada}}, \bibinfo {author} {\bibfnamefont {S.}~\bibnamefont {Okuma}}, \bibinfo {author} {\bibfnamefont {M.}~\bibnamefont {Ishikawa}}, \bibinfo {author} {\bibfnamefont {T.}~\bibnamefont {Takabatake}}, \bibinfo {author} {\bibfnamefont {Y.}~\bibnamefont {Nakazawa}}, \bibinfo {author} {\bibfnamefont {Y.}~\bibnamefont {Kuno}}, \bibinfo {author} {\bibfnamefont {R.}~\bibnamefont {Keitel}}, \bibinfo {author} {\bibfnamefont {J.~H.}\ \bibnamefont {Brewer}}, \bibinfo {author} {\bibfnamefont {T.~M.}\ \bibnamefont {Riseman}}, \bibinfo {author} {\bibfnamefont {D.~L.}\ \bibnamefont {Williams}}, \bibinfo {author} {\bibfnamefont {Y.}~\bibnamefont {Watanabe}}, \bibinfo {author} {\bibfnamefont {T.}~\bibnamefont {Yamazaki}}, \bibinfo {author} {\bibfnamefont {K.}~\bibnamefont {Nishiyama}}, \bibinfo {author} {\bibfnamefont
  {K.}~\bibnamefont {Nagamine}}, \bibinfo {author} {\bibfnamefont {E.~J.}\ \bibnamefont {Ansaldo}},\ and\ \bibinfo {author} {\bibfnamefont {E.}~\bibnamefont {Torikai}},\ }\href {https://doi.org/10.1143/JJAP.26.L1856} {\bibfield  {journal} {\bibinfo  {journal} {Jpn. J. Appl. Phys.}\ }\textbf {\bibinfo {volume} {26}},\ \bibinfo {pages} {L1856} (\bibinfo {year} {1987})}\BibitemShut {NoStop}%
\bibitem [{\citenamefont {Pümpin}\ \emph {et~al.}(1988)\citenamefont {Pümpin}, \citenamefont {Keller}, \citenamefont {Kündig}, \citenamefont {Odermatt}, \citenamefont {Patterson}, \citenamefont {Schneider}, \citenamefont {Simmler}, \citenamefont {Connell}, \citenamefont {Müller}, \citenamefont {Bednorz}, \citenamefont {Blazey}, \citenamefont {Morgenstern}, \citenamefont {Rossel},\ and\ \citenamefont {Savić}}]{pumpin_internal_1988}%
  \BibitemOpen
  \bibfield  {author} {\bibinfo {author} {\bibfnamefont {B.}~\bibnamefont {Pümpin}}, \bibinfo {author} {\bibfnamefont {H.}~\bibnamefont {Keller}}, \bibinfo {author} {\bibfnamefont {W.}~\bibnamefont {Kündig}}, \bibinfo {author} {\bibfnamefont {W.}~\bibnamefont {Odermatt}}, \bibinfo {author} {\bibfnamefont {B.~D.}\ \bibnamefont {Patterson}}, \bibinfo {author} {\bibfnamefont {J.~W.}\ \bibnamefont {Schneider}}, \bibinfo {author} {\bibfnamefont {H.}~\bibnamefont {Simmler}}, \bibinfo {author} {\bibfnamefont {S.}~\bibnamefont {Connell}}, \bibinfo {author} {\bibfnamefont {K.~A.}\ \bibnamefont {Müller}}, \bibinfo {author} {\bibfnamefont {J.~G.}\ \bibnamefont {Bednorz}}, \bibinfo {author} {\bibfnamefont {K.~W.}\ \bibnamefont {Blazey}}, \bibinfo {author} {\bibfnamefont {I.}~\bibnamefont {Morgenstern}}, \bibinfo {author} {\bibfnamefont {C.}~\bibnamefont {Rossel}},\ and\ \bibinfo {author} {\bibfnamefont {I.~M.}\ \bibnamefont {Savić}},\ }\href {https://doi.org/10.1007/BF01312133} {\bibfield  {journal} {\bibinfo
  {journal} {Z. Physik B - Condensed Matter}\ }\textbf {\bibinfo {volume} {72}},\ \bibinfo {pages} {175} (\bibinfo {year} {1988})}\BibitemShut {NoStop}%
\bibitem [{\citenamefont {Brewer}\ \emph {et~al.}(1988)\citenamefont {Brewer}, \citenamefont {Ansaldo}, \citenamefont {Carolan}, \citenamefont {Chaklader}, \citenamefont {Hardy}, \citenamefont {Harshman}, \citenamefont {Hayden}, \citenamefont {Ishikawa}, \citenamefont {Kaplan}, \citenamefont {Keitel}, \citenamefont {Kempton}, \citenamefont {Kiefl}, \citenamefont {Kossler}, \citenamefont {Kreitzman}, \citenamefont {Kulpa}, \citenamefont {Kuno}, \citenamefont {Luke}, \citenamefont {Miyatake}, \citenamefont {Nagamine}, \citenamefont {Nakazawa}, \citenamefont {Nishida}, \citenamefont {Nishiyama}, \citenamefont {Ohkuma}, \citenamefont {Riseman}, \citenamefont {Roehmer}, \citenamefont {Schleger}, \citenamefont {Shimada}, \citenamefont {Stronach}, \citenamefont {Takabatake}, \citenamefont {Uemura}, \citenamefont {Watanabe}, \citenamefont {Williams}, \citenamefont {Yamazaki},\ and\ \citenamefont {Yang}}]{brewer_antiferromagnetism_1988}%
  \BibitemOpen
  \bibfield  {author} {\bibinfo {author} {\bibfnamefont {J.~H.}\ \bibnamefont {Brewer}}, \bibinfo {author} {\bibfnamefont {E.~J.}\ \bibnamefont {Ansaldo}}, \bibinfo {author} {\bibfnamefont {J.~F.}\ \bibnamefont {Carolan}}, \bibinfo {author} {\bibfnamefont {A.~C.~D.}\ \bibnamefont {Chaklader}}, \bibinfo {author} {\bibfnamefont {W.~N.}\ \bibnamefont {Hardy}}, \bibinfo {author} {\bibfnamefont {D.~R.}\ \bibnamefont {Harshman}}, \bibinfo {author} {\bibfnamefont {M.~E.}\ \bibnamefont {Hayden}}, \bibinfo {author} {\bibfnamefont {M.}~\bibnamefont {Ishikawa}}, \bibinfo {author} {\bibfnamefont {N.}~\bibnamefont {Kaplan}}, \bibinfo {author} {\bibfnamefont {R.}~\bibnamefont {Keitel}}, \bibinfo {author} {\bibfnamefont {J.}~\bibnamefont {Kempton}}, \bibinfo {author} {\bibfnamefont {R.~F.}\ \bibnamefont {Kiefl}}, \bibinfo {author} {\bibfnamefont {W.~J.}\ \bibnamefont {Kossler}}, \bibinfo {author} {\bibfnamefont {S.~R.}\ \bibnamefont {Kreitzman}}, \bibinfo {author} {\bibfnamefont {A.}~\bibnamefont {Kulpa}}, \bibinfo {author}
  {\bibfnamefont {Y.}~\bibnamefont {Kuno}}, \bibinfo {author} {\bibfnamefont {G.~M.}\ \bibnamefont {Luke}}, \bibinfo {author} {\bibfnamefont {H.}~\bibnamefont {Miyatake}}, \bibinfo {author} {\bibfnamefont {K.}~\bibnamefont {Nagamine}}, \bibinfo {author} {\bibfnamefont {Y.}~\bibnamefont {Nakazawa}}, \bibinfo {author} {\bibfnamefont {N.}~\bibnamefont {Nishida}}, \bibinfo {author} {\bibfnamefont {K.}~\bibnamefont {Nishiyama}}, \bibinfo {author} {\bibfnamefont {S.}~\bibnamefont {Ohkuma}}, \bibinfo {author} {\bibfnamefont {T.~M.}\ \bibnamefont {Riseman}}, \bibinfo {author} {\bibfnamefont {G.}~\bibnamefont {Roehmer}}, \bibinfo {author} {\bibfnamefont {P.}~\bibnamefont {Schleger}}, \bibinfo {author} {\bibfnamefont {D.}~\bibnamefont {Shimada}}, \bibinfo {author} {\bibfnamefont {C.~E.}\ \bibnamefont {Stronach}}, \bibinfo {author} {\bibfnamefont {T.}~\bibnamefont {Takabatake}}, \bibinfo {author} {\bibfnamefont {Y.~J.}\ \bibnamefont {Uemura}}, \bibinfo {author} {\bibfnamefont {Y.}~\bibnamefont {Watanabe}}, \bibinfo
  {author} {\bibfnamefont {D.~L.}\ \bibnamefont {Williams}}, \bibinfo {author} {\bibfnamefont {T.}~\bibnamefont {Yamazaki}},\ and\ \bibinfo {author} {\bibfnamefont {B.}~\bibnamefont {Yang}},\ }\href {https://doi.org/10.1103/PhysRevLett.60.1073} {\bibfield  {journal} {\bibinfo  {journal} {Phys. Rev. Lett.}\ }\textbf {\bibinfo {volume} {60}},\ \bibinfo {pages} {1073} (\bibinfo {year} {1988})}\BibitemShut {NoStop}%
\bibitem [{\citenamefont {Higemoto}\ \emph {et~al.}(2014)\citenamefont {Higemoto}, \citenamefont {Koda}, \citenamefont {Kadono}, \citenamefont {Yoshida},\ and\ \citenamefont {Onuki}}]{higemoto_investigation_2014}%
  \BibitemOpen
  \bibfield  {author} {\bibinfo {author} {\bibfnamefont {W.}~\bibnamefont {Higemoto}}, \bibinfo {author} {\bibfnamefont {A.}~\bibnamefont {Koda}}, \bibinfo {author} {\bibfnamefont {R.}~\bibnamefont {Kadono}}, \bibinfo {author} {\bibfnamefont {Y.}~\bibnamefont {Yoshida}},\ and\ \bibinfo {author} {\bibfnamefont {Y.}~\bibnamefont {Onuki}},\ }\href {https://doi.org/10.7566/JPSCP.2.010202} {\bibfield  {journal} {\bibinfo  {journal} {Proc. Int. Symp. Science Explored by Ultra Slow Muon (USM2013)}\ }\bibinfo {series} {{JPS} {Conference} {Proceedings}},\ \textbf {\bibinfo {volume} {2}},\ \bibinfo {pages} {010202} (\bibinfo {year} {2014})}\BibitemShut {NoStop}%
\bibitem [{\citenamefont {Fittipaldi}\ \emph {et~al.}(2021)\citenamefont {Fittipaldi}, \citenamefont {Hartmann}, \citenamefont {Mercaldo}, \citenamefont {Komori}, \citenamefont {Bjørlig}, \citenamefont {Kyung}, \citenamefont {Yasui}, \citenamefont {Miyoshi}, \citenamefont {Olde~Olthof}, \citenamefont {Palomares~Garcia}, \citenamefont {Granata}, \citenamefont {Keren}, \citenamefont {Higemoto}, \citenamefont {Suter}, \citenamefont {Prokscha}, \citenamefont {Romano}, \citenamefont {Noce}, \citenamefont {Kim}, \citenamefont {Maeno}, \citenamefont {Scheer}, \citenamefont {Kalisky}, \citenamefont {Robinson}, \citenamefont {Cuoco}, \citenamefont {Salman}, \citenamefont {Vecchione},\ and\ \citenamefont {Di~Bernardo}}]{fittipaldi_unveiling_2021}%
  \BibitemOpen
  \bibfield  {author} {\bibinfo {author} {\bibfnamefont {R.}~\bibnamefont {Fittipaldi}}, \bibinfo {author} {\bibfnamefont {R.}~\bibnamefont {Hartmann}}, \bibinfo {author} {\bibfnamefont {M.~T.}\ \bibnamefont {Mercaldo}}, \bibinfo {author} {\bibfnamefont {S.}~\bibnamefont {Komori}}, \bibinfo {author} {\bibfnamefont {A.}~\bibnamefont {Bjørlig}}, \bibinfo {author} {\bibfnamefont {W.}~\bibnamefont {Kyung}}, \bibinfo {author} {\bibfnamefont {Y.}~\bibnamefont {Yasui}}, \bibinfo {author} {\bibfnamefont {T.}~\bibnamefont {Miyoshi}}, \bibinfo {author} {\bibfnamefont {L.~A.~B.}\ \bibnamefont {Olde~Olthof}}, \bibinfo {author} {\bibfnamefont {C.~M.}\ \bibnamefont {Palomares~Garcia}}, \bibinfo {author} {\bibfnamefont {V.}~\bibnamefont {Granata}}, \bibinfo {author} {\bibfnamefont {I.}~\bibnamefont {Keren}}, \bibinfo {author} {\bibfnamefont {W.}~\bibnamefont {Higemoto}}, \bibinfo {author} {\bibfnamefont {A.}~\bibnamefont {Suter}}, \bibinfo {author} {\bibfnamefont {T.}~\bibnamefont {Prokscha}}, \bibinfo {author}
  {\bibfnamefont {A.}~\bibnamefont {Romano}}, \bibinfo {author} {\bibfnamefont {C.}~\bibnamefont {Noce}}, \bibinfo {author} {\bibfnamefont {C.}~\bibnamefont {Kim}}, \bibinfo {author} {\bibfnamefont {Y.}~\bibnamefont {Maeno}}, \bibinfo {author} {\bibfnamefont {E.}~\bibnamefont {Scheer}}, \bibinfo {author} {\bibfnamefont {B.}~\bibnamefont {Kalisky}}, \bibinfo {author} {\bibfnamefont {J.~W.~A.}\ \bibnamefont {Robinson}}, \bibinfo {author} {\bibfnamefont {M.}~\bibnamefont {Cuoco}}, \bibinfo {author} {\bibfnamefont {Z.}~\bibnamefont {Salman}}, \bibinfo {author} {\bibfnamefont {A.}~\bibnamefont {Vecchione}},\ and\ \bibinfo {author} {\bibfnamefont {A.}~\bibnamefont {Di~Bernardo}},\ }\href {https://doi.org/10.1038/s41467-021-26020-5} {\bibfield  {journal} {\bibinfo  {journal} {Nat. Commun.}\ }\textbf {\bibinfo {volume} {12}},\ \bibinfo {pages} {5792} (\bibinfo {year} {2021})}\BibitemShut {NoStop}%
\bibitem [{sup()}]{supplement_2025}%
  \BibitemOpen
  \href@noop {} {\bibinfo  {journal} {See Supplemental Material available at [URL], which includes Ref. [2, 5-10, 30, 31, 46-63], for the comparison of various methods of probing magnetism of a superconductor, the details of sample growth and characterization, the FLAME spectrometer and its spectral analysis, and the DC magnetization measurements}\ }\BibitemShut {NoStop}%
\bibitem [{\citenamefont {Jang}\ \emph {et~al.}(2011{\natexlab{a}})\citenamefont {Jang}, \citenamefont {Budakian},\ and\ \citenamefont {Maeno}}]{jang_phase-locked_2011}%
  \BibitemOpen
\bibfield  {journal} {  }\bibfield  {author} {\bibinfo {author} {\bibfnamefont {J.}~\bibnamefont {Jang}}, \bibinfo {author} {\bibfnamefont {R.}~\bibnamefont {Budakian}},\ and\ \bibinfo {author} {\bibfnamefont {Y.}~\bibnamefont {Maeno}},\ }\href {https://doi.org/10.1063/1.3572026} {\bibfield  {journal} {\bibinfo  {journal} {Appl. Phys. Lett.}\ }\textbf {\bibinfo {volume} {98}},\ \bibinfo {pages} {132510} (\bibinfo {year} {2011}{\natexlab{a}})}\BibitemShut {NoStop}%
\bibitem [{\citenamefont {Jang}\ \emph {et~al.}(2011{\natexlab{b}})\citenamefont {Jang}, \citenamefont {Ferguson}, \citenamefont {Vakaryuk}, \citenamefont {Budakian}, \citenamefont {Chung}, \citenamefont {Goldbart},\ and\ \citenamefont {Maeno}}]{jang_observation_2011}%
  \BibitemOpen
  \bibfield  {author} {\bibinfo {author} {\bibfnamefont {J.}~\bibnamefont {Jang}}, \bibinfo {author} {\bibfnamefont {D.~G.}\ \bibnamefont {Ferguson}}, \bibinfo {author} {\bibfnamefont {V.}~\bibnamefont {Vakaryuk}}, \bibinfo {author} {\bibfnamefont {R.}~\bibnamefont {Budakian}}, \bibinfo {author} {\bibfnamefont {S.~B.}\ \bibnamefont {Chung}}, \bibinfo {author} {\bibfnamefont {P.~M.}\ \bibnamefont {Goldbart}},\ and\ \bibinfo {author} {\bibfnamefont {Y.}~\bibnamefont {Maeno}},\ }\href {https://doi.org/10.1126/science.1193839} {\bibfield  {journal} {\bibinfo  {journal} {Science}\ }\textbf {\bibinfo {volume} {331}},\ \bibinfo {pages} {186} (\bibinfo {year} {2011}{\natexlab{b}})}\BibitemShut {NoStop}%
\bibitem [{\citenamefont {Yonezawa}\ \emph {et~al.}(2015)\citenamefont {Yonezawa}, \citenamefont {Higuchi}, \citenamefont {Sugimoto}, \citenamefont {Sow},\ and\ \citenamefont {Maeno}}]{yonezawa_compact_2015}%
  \BibitemOpen
  \bibfield  {author} {\bibinfo {author} {\bibfnamefont {S.}~\bibnamefont {Yonezawa}}, \bibinfo {author} {\bibfnamefont {T.}~\bibnamefont {Higuchi}}, \bibinfo {author} {\bibfnamefont {Y.}~\bibnamefont {Sugimoto}}, \bibinfo {author} {\bibfnamefont {C.}~\bibnamefont {Sow}},\ and\ \bibinfo {author} {\bibfnamefont {Y.}~\bibnamefont {Maeno}},\ }\href {https://doi.org/10.1063/1.4929871} {\bibfield  {journal} {\bibinfo  {journal} {Rev. Sci. Instrum.}\ }\textbf {\bibinfo {volume} {86}},\ \bibinfo {pages} {093903} (\bibinfo {year} {2015})}\BibitemShut {NoStop}%
\bibitem [{\citenamefont {Tenya}\ \emph {et~al.}(2006)\citenamefont {Tenya}, \citenamefont {Yasuda}, \citenamefont {Yokoyama}, \citenamefont {Amitsuka}, \citenamefont {Deguchi},\ and\ \citenamefont {Maeno}}]{tenya_unusual_2006}%
  \BibitemOpen
  \bibfield  {author} {\bibinfo {author} {\bibfnamefont {K.}~\bibnamefont {Tenya}}, \bibinfo {author} {\bibfnamefont {S.}~\bibnamefont {Yasuda}}, \bibinfo {author} {\bibfnamefont {M.}~\bibnamefont {Yokoyama}}, \bibinfo {author} {\bibfnamefont {H.}~\bibnamefont {Amitsuka}}, \bibinfo {author} {\bibfnamefont {K.}~\bibnamefont {Deguchi}},\ and\ \bibinfo {author} {\bibfnamefont {Y.}~\bibnamefont {Maeno}},\ }\href {https://doi.org/10.1143/JPSJ.75.023702} {\bibfield  {journal} {\bibinfo  {journal} {J. Phys. Soc. Jpn.}\ }\textbf {\bibinfo {volume} {75}},\ \bibinfo {pages} {023702} (\bibinfo {year} {2006})}\BibitemShut {NoStop}%
\bibitem [{\citenamefont {Kittaka}\ \emph {et~al.}(2014)\citenamefont {Kittaka}, \citenamefont {Kasahara}, \citenamefont {Sakakibara}, \citenamefont {Shibata}, \citenamefont {Yonezawa}, \citenamefont {Maeno}, \citenamefont {Tenya},\ and\ \citenamefont {Machida}}]{kittaka_sharp_2014}%
  \BibitemOpen
  \bibfield  {author} {\bibinfo {author} {\bibfnamefont {S.}~\bibnamefont {Kittaka}}, \bibinfo {author} {\bibfnamefont {A.}~\bibnamefont {Kasahara}}, \bibinfo {author} {\bibfnamefont {T.}~\bibnamefont {Sakakibara}}, \bibinfo {author} {\bibfnamefont {D.}~\bibnamefont {Shibata}}, \bibinfo {author} {\bibfnamefont {S.}~\bibnamefont {Yonezawa}}, \bibinfo {author} {\bibfnamefont {Y.}~\bibnamefont {Maeno}}, \bibinfo {author} {\bibfnamefont {K.}~\bibnamefont {Tenya}},\ and\ \bibinfo {author} {\bibfnamefont {K.}~\bibnamefont {Machida}},\ }\href {https://doi.org/10.1103/PhysRevB.90.220502} {\bibfield  {journal} {\bibinfo  {journal} {Phys. Rev. B}\ }\textbf {\bibinfo {volume} {90}},\ \bibinfo {pages} {220502} (\bibinfo {year} {2014})}\BibitemShut {NoStop}%
\bibitem [{\citenamefont {Shimizu}\ \emph {et~al.}(2021)\citenamefont {Shimizu}, \citenamefont {Okumura}, \citenamefont {Kato},\ and\ \citenamefont {Motome}}]{shimizu_spin_2021}%
  \BibitemOpen
  \bibfield  {author} {\bibinfo {author} {\bibfnamefont {K.}~\bibnamefont {Shimizu}}, \bibinfo {author} {\bibfnamefont {S.}~\bibnamefont {Okumura}}, \bibinfo {author} {\bibfnamefont {Y.}~\bibnamefont {Kato}},\ and\ \bibinfo {author} {\bibfnamefont {Y.}~\bibnamefont {Motome}},\ }\href {https://doi.org/10.1103/PhysRevB.103.184421} {\bibfield  {journal} {\bibinfo  {journal} {Phys. Rev. B}\ }\textbf {\bibinfo {volume} {103}},\ \bibinfo {pages} {184421} (\bibinfo {year} {2021})}\BibitemShut {NoStop}%
\bibitem [{\citenamefont {Shull}\ and\ \citenamefont {Wedgwood}(1966)}]{shull_neutron-diffraction_1966}%
  \BibitemOpen
  \bibfield  {author} {\bibinfo {author} {\bibfnamefont {C.~G.}\ \bibnamefont {Shull}}\ and\ \bibinfo {author} {\bibfnamefont {F.~A.}\ \bibnamefont {Wedgwood}},\ }\href {https://doi.org/10.1103/PhysRevLett.16.513} {\bibfield  {journal} {\bibinfo  {journal} {Phys. Rev. Lett.}\ }\textbf {\bibinfo {volume} {16}},\ \bibinfo {pages} {513} (\bibinfo {year} {1966})}\BibitemShut {NoStop}%
\bibitem [{\citenamefont {Mukuda}\ \emph {et~al.}(1999)\citenamefont {Mukuda}, \citenamefont {Ishida}, \citenamefont {Kitaoka}, \citenamefont {Asayama}, \citenamefont {Kanno},\ and\ \citenamefont {Takano}}]{mukuda_spin_1999}%
  \BibitemOpen
  \bibfield  {author} {\bibinfo {author} {\bibfnamefont {H.}~\bibnamefont {Mukuda}}, \bibinfo {author} {\bibfnamefont {K.}~\bibnamefont {Ishida}}, \bibinfo {author} {\bibfnamefont {Y.}~\bibnamefont {Kitaoka}}, \bibinfo {author} {\bibfnamefont {K.}~\bibnamefont {Asayama}}, \bibinfo {author} {\bibfnamefont {R.}~\bibnamefont {Kanno}},\ and\ \bibinfo {author} {\bibfnamefont {M.}~\bibnamefont {Takano}},\ }\href {https://doi.org/10.1103/PhysRevB.60.12279} {\bibfield  {journal} {\bibinfo  {journal} {Phys. Rev. B}\ }\textbf {\bibinfo {volume} {60}},\ \bibinfo {pages} {12279} (\bibinfo {year} {1999})}\BibitemShut {NoStop}%
\bibitem [{\citenamefont {Ghosh}\ \emph {et~al.}(2020)\citenamefont {Ghosh}, \citenamefont {Smidman}, \citenamefont {Shang}, \citenamefont {Annett}, \citenamefont {Hillier}, \citenamefont {Quintanilla},\ and\ \citenamefont {Yuan}}]{ghosh_recent_2020}%
  \BibitemOpen
  \bibfield  {author} {\bibinfo {author} {\bibfnamefont {S.~K.}\ \bibnamefont {Ghosh}}, \bibinfo {author} {\bibfnamefont {M.}~\bibnamefont {Smidman}}, \bibinfo {author} {\bibfnamefont {T.}~\bibnamefont {Shang}}, \bibinfo {author} {\bibfnamefont {J.~F.}\ \bibnamefont {Annett}}, \bibinfo {author} {\bibfnamefont {A.~D.}\ \bibnamefont {Hillier}}, \bibinfo {author} {\bibfnamefont {J.}~\bibnamefont {Quintanilla}},\ and\ \bibinfo {author} {\bibfnamefont {H.}~\bibnamefont {Yuan}},\ }\href {https://doi.org/10.1088/1361-648X/abaa06} {\bibfield  {journal} {\bibinfo  {journal} {J. Phys.: Condens. Matter}\ }\textbf {\bibinfo {volume} {33}},\ \bibinfo {pages} {033001} (\bibinfo {year} {2020})}\BibitemShut {NoStop}%
\bibitem [{\citenamefont {Xia}\ \emph {et~al.}(2006{\natexlab{a}})\citenamefont {Xia}, \citenamefont {Maeno}, \citenamefont {Beyersdorf}, \citenamefont {Fejer},\ and\ \citenamefont {Kapitulnik}}]{xia_high_2006}%
  \BibitemOpen
  \bibfield  {author} {\bibinfo {author} {\bibfnamefont {J.}~\bibnamefont {Xia}}, \bibinfo {author} {\bibfnamefont {Y.}~\bibnamefont {Maeno}}, \bibinfo {author} {\bibfnamefont {P.~T.}\ \bibnamefont {Beyersdorf}}, \bibinfo {author} {\bibfnamefont {M.~M.}\ \bibnamefont {Fejer}},\ and\ \bibinfo {author} {\bibfnamefont {A.}~\bibnamefont {Kapitulnik}},\ }\href {https://doi.org/10.1103/PhysRevLett.97.167002} {\bibfield  {journal} {\bibinfo  {journal} {Phys. Rev. Lett.}\ }\textbf {\bibinfo {volume} {97}},\ \bibinfo {pages} {167002} (\bibinfo {year} {2006}{\natexlab{a}})}\BibitemShut {NoStop}%
\bibitem [{\citenamefont {Xia}\ \emph {et~al.}(2006{\natexlab{b}})\citenamefont {Xia}, \citenamefont {Beyersdorf}, \citenamefont {Fejer},\ and\ \citenamefont {Kapitulnik}}]{xia_modified_2006}%
  \BibitemOpen
  \bibfield  {author} {\bibinfo {author} {\bibfnamefont {J.}~\bibnamefont {Xia}}, \bibinfo {author} {\bibfnamefont {P.~T.}\ \bibnamefont {Beyersdorf}}, \bibinfo {author} {\bibfnamefont {M.~M.}\ \bibnamefont {Fejer}},\ and\ \bibinfo {author} {\bibfnamefont {A.}~\bibnamefont {Kapitulnik}},\ }\href {https://doi.org/10.1063/1.2336620} {\bibfield  {journal} {\bibinfo  {journal} {Appl. Phys. Lett.}\ }\textbf {\bibinfo {volume} {89}},\ \bibinfo {pages} {062508} (\bibinfo {year} {2006}{\natexlab{b}})}\BibitemShut {NoStop}%
\bibitem [{\citenamefont {Bobowski}\ \emph {et~al.}(2019)\citenamefont {Bobowski}, \citenamefont {Kikugawa}, \citenamefont {Miyoshi}, \citenamefont {Suwa}, \citenamefont {Xu}, \citenamefont {Yonezawa}, \citenamefont {Sokolov}, \citenamefont {Mackenzie},\ and\ \citenamefont {Maeno}}]{bobowski_improved_2019}%
  \BibitemOpen
  \bibfield  {author} {\bibinfo {author} {\bibfnamefont {J.~S.}\ \bibnamefont {Bobowski}}, \bibinfo {author} {\bibfnamefont {N.}~\bibnamefont {Kikugawa}}, \bibinfo {author} {\bibfnamefont {T.}~\bibnamefont {Miyoshi}}, \bibinfo {author} {\bibfnamefont {H.}~\bibnamefont {Suwa}}, \bibinfo {author} {\bibfnamefont {H.-s.}\ \bibnamefont {Xu}}, \bibinfo {author} {\bibfnamefont {S.}~\bibnamefont {Yonezawa}}, \bibinfo {author} {\bibfnamefont {D.~A.}\ \bibnamefont {Sokolov}}, \bibinfo {author} {\bibfnamefont {A.~P.}\ \bibnamefont {Mackenzie}},\ and\ \bibinfo {author} {\bibfnamefont {Y.}~\bibnamefont {Maeno}},\ }\href {https://doi.org/10.3390/condmat4010006} {\bibfield  {journal} {\bibinfo  {journal} {Condens. Matter}\ }\textbf {\bibinfo {volume} {4}},\ \bibinfo {pages} {6} (\bibinfo {year} {2019})}\BibitemShut {NoStop}%
\bibitem [{\citenamefont {Sato}\ and\ \citenamefont {Ishii}(1989)}]{sato_simple_1989}%
  \BibitemOpen
  \bibfield  {author} {\bibinfo {author} {\bibfnamefont {M.}~\bibnamefont {Sato}}\ and\ \bibinfo {author} {\bibfnamefont {Y.}~\bibnamefont {Ishii}},\ }\href {https://doi.org/10.1063/1.343481} {\bibfield  {journal} {\bibinfo  {journal} {J. Appl. Phys.}\ }\textbf {\bibinfo {volume} {66}},\ \bibinfo {pages} {983} (\bibinfo {year} {1989})}\BibitemShut {NoStop}%
\bibitem [{\citenamefont {Huddart}\ \emph {et~al.}(2021)\citenamefont {Huddart}, \citenamefont {Onuorah}, \citenamefont {Isah}, \citenamefont {Bonfà}, \citenamefont {Blundell}, \citenamefont {Clark}, \citenamefont {De~Renzi},\ and\ \citenamefont {Lancaster}}]{huddart_intrinsic_2021}%
  \BibitemOpen
  \bibfield  {author} {\bibinfo {author} {\bibfnamefont {B.}~\bibnamefont {Huddart}}, \bibinfo {author} {\bibfnamefont {I.}~\bibnamefont {Onuorah}}, \bibinfo {author} {\bibfnamefont {M.}~\bibnamefont {Isah}}, \bibinfo {author} {\bibfnamefont {P.}~\bibnamefont {Bonfà}}, \bibinfo {author} {\bibfnamefont {S.}~\bibnamefont {Blundell}}, \bibinfo {author} {\bibfnamefont {S.}~\bibnamefont {Clark}}, \bibinfo {author} {\bibfnamefont {R.}~\bibnamefont {De~Renzi}},\ and\ \bibinfo {author} {\bibfnamefont {T.}~\bibnamefont {Lancaster}},\ }\href {https://doi.org/10.1103/PhysRevLett.127.237002} {\bibfield  {journal} {\bibinfo  {journal} {Phys. Rev. Lett.}\ }\textbf {\bibinfo {volume} {127}},\ \bibinfo {pages} {237002} (\bibinfo {year} {2021})}\BibitemShut {NoStop}%
\bibitem [{\citenamefont {Blundell}\ and\ \citenamefont {Lancaster}(2023)}]{blundell_dft_2023}%
  \BibitemOpen
  \bibfield  {author} {\bibinfo {author} {\bibfnamefont {S.~J.}\ \bibnamefont {Blundell}}\ and\ \bibinfo {author} {\bibfnamefont {T.}~\bibnamefont {Lancaster}},\ }\href {https://doi.org/10.1063/5.0149080} {\bibfield  {journal} {\bibinfo  {journal} {Appl. Phys. Rev.}\ }\textbf {\bibinfo {volume} {10}},\ \bibinfo {pages} {021316} (\bibinfo {year} {2023})}\BibitemShut {NoStop}%
\bibitem [{\citenamefont {Hotz}\ \emph {et~al.}(2023)\citenamefont {Hotz}, \citenamefont {Arh}, \citenamefont {Guguchia}, \citenamefont {Das}, \citenamefont {Wang}, \citenamefont {Gomilsek}, \citenamefont {Zorko},\ and\ \citenamefont {Luetkens}}]{hotz_experimental_2023}%
  \BibitemOpen
  \bibfield  {author} {\bibinfo {author} {\bibfnamefont {F.}~\bibnamefont {Hotz}}, \bibinfo {author} {\bibfnamefont {T.}~\bibnamefont {Arh}}, \bibinfo {author} {\bibfnamefont {Z.}~\bibnamefont {Guguchia}}, \bibinfo {author} {\bibfnamefont {D.}~\bibnamefont {Das}}, \bibinfo {author} {\bibfnamefont {C.}~\bibnamefont {Wang}}, \bibinfo {author} {\bibfnamefont {M.}~\bibnamefont {Gomilsek}}, \bibinfo {author} {\bibfnamefont {A.}~\bibnamefont {Zorko}},\ and\ \bibinfo {author} {\bibfnamefont {H.}~\bibnamefont {Luetkens}},\ }\href {https://doi.org/10.1088/1742-6596/2462/1/012041} {\bibfield  {journal} {\bibinfo  {journal} {J. Phys.: Conf. Ser.}\ }\textbf {\bibinfo {volume} {2462}},\ \bibinfo {pages} {012041} (\bibinfo {year} {2023})}\BibitemShut {NoStop}%
\bibitem [{\citenamefont {Kittaka}\ \emph {et~al.}(2009)\citenamefont {Kittaka}, \citenamefont {Nakamura}, \citenamefont {Aono}, \citenamefont {Yonezawa}, \citenamefont {Ishida},\ and\ \citenamefont {Maeno}}]{kittaka_angular_2009}%
  \BibitemOpen
  \bibfield  {author} {\bibinfo {author} {\bibfnamefont {S.}~\bibnamefont {Kittaka}}, \bibinfo {author} {\bibfnamefont {T.}~\bibnamefont {Nakamura}}, \bibinfo {author} {\bibfnamefont {Y.}~\bibnamefont {Aono}}, \bibinfo {author} {\bibfnamefont {S.}~\bibnamefont {Yonezawa}}, \bibinfo {author} {\bibfnamefont {K.}~\bibnamefont {Ishida}},\ and\ \bibinfo {author} {\bibfnamefont {Y.}~\bibnamefont {Maeno}},\ }\href {https://doi.org/10.1103/PhysRevB.80.174514} {\bibfield  {journal} {\bibinfo  {journal} {Phys. Rev. B}\ }\textbf {\bibinfo {volume} {80}},\ \bibinfo {pages} {174514} (\bibinfo {year} {2009})}\BibitemShut {NoStop}%
\bibitem [{\citenamefont {Maeno}\ \emph {et~al.}(1997)\citenamefont {Maeno}, \citenamefont {Yoshida}, \citenamefont {Hashimoto}, \citenamefont {Nishizaki}, \citenamefont {Ikeda}, \citenamefont {Nohara}, \citenamefont {Fujita}, \citenamefont {Mackenzie}, \citenamefont {Hussey}, \citenamefont {Bednorz},\ and\ \citenamefont {Lichtenberg}}]{maeno_two-dimensional_1997}%
  \BibitemOpen
  \bibfield  {author} {\bibinfo {author} {\bibfnamefont {Y.}~\bibnamefont {Maeno}}, \bibinfo {author} {\bibfnamefont {K.}~\bibnamefont {Yoshida}}, \bibinfo {author} {\bibfnamefont {H.}~\bibnamefont {Hashimoto}}, \bibinfo {author} {\bibfnamefont {S.}~\bibnamefont {Nishizaki}}, \bibinfo {author} {\bibfnamefont {S.-i.}\ \bibnamefont {Ikeda}}, \bibinfo {author} {\bibfnamefont {M.}~\bibnamefont {Nohara}}, \bibinfo {author} {\bibfnamefont {T.}~\bibnamefont {Fujita}}, \bibinfo {author} {\bibfnamefont {A.}~\bibnamefont {Mackenzie}}, \bibinfo {author} {\bibfnamefont {N.}~\bibnamefont {Hussey}}, \bibinfo {author} {\bibfnamefont {J.}~\bibnamefont {Bednorz}},\ and\ \bibinfo {author} {\bibfnamefont {F.}~\bibnamefont {Lichtenberg}},\ }\href {https://doi.org/10.1143/JPSJ.66.1405} {\bibfield  {journal} {\bibinfo  {journal} {J. Phys. Soc. Jpn.}\ }\textbf {\bibinfo {volume} {66}},\ \bibinfo {pages} {1405} (\bibinfo {year} {1997})}\BibitemShut {NoStop}%
\bibitem [{muS()}]{muSR_2025}%
  \BibitemOpen
  \href@noop {} {\bibinfo  {journal} {http://musruser.psi.ch/cgi-bin/SearchDB.cgi}\ }\BibitemShut {NoStop}%
\bibitem [{\citenamefont {Amano}\ \emph {et~al.}(2015)\citenamefont {Amano}, \citenamefont {Ishihara}, \citenamefont {Ichioka}, \citenamefont {Nakai},\ and\ \citenamefont {Machida}}]{amano_pauli_2015}%
  \BibitemOpen
\bibfield  {journal} {  }\bibfield  {author} {\bibinfo {author} {\bibfnamefont {Y.}~\bibnamefont {Amano}}, \bibinfo {author} {\bibfnamefont {M.}~\bibnamefont {Ishihara}}, \bibinfo {author} {\bibfnamefont {M.}~\bibnamefont {Ichioka}}, \bibinfo {author} {\bibfnamefont {N.}~\bibnamefont {Nakai}},\ and\ \bibinfo {author} {\bibfnamefont {K.}~\bibnamefont {Machida}},\ }\href {https://doi.org/10.1103/PhysRevB.91.144513} {\bibfield  {journal} {\bibinfo  {journal} {Phys. Rev. B}\ }\textbf {\bibinfo {volume} {91}},\ \bibinfo {pages} {144513} (\bibinfo {year} {2015})}\BibitemShut {NoStop}%
\bibitem [{\citenamefont {Udagawa}\ and\ \citenamefont {Yanase}(2010)}]{udagawa_correlation_2010}%
  \BibitemOpen
  \bibfield  {author} {\bibinfo {author} {\bibfnamefont {M.}~\bibnamefont {Udagawa}}\ and\ \bibinfo {author} {\bibfnamefont {Y.}~\bibnamefont {Yanase}},\ }\href {https://doi.org/10.1016/j.physc.2009.11.115} {\bibfield  {journal} {\bibinfo  {journal} {Physica C: Superconductivity and its Applications}\ }\textbf {\bibinfo {volume} {470}},\ \bibinfo {pages} {S697} (\bibinfo {year} {2010})}\BibitemShut {NoStop}%
\bibitem [{\citenamefont {Mackenzie}\ and\ \citenamefont {Maeno}(2003)}]{mackenzie_superconductivity_2003}%
  \BibitemOpen
  \bibfield  {author} {\bibinfo {author} {\bibfnamefont {A.~P.}\ \bibnamefont {Mackenzie}}\ and\ \bibinfo {author} {\bibfnamefont {Y.}~\bibnamefont {Maeno}},\ }\href {https://doi.org/10.1103/RevModPhys.75.657} {\bibfield  {journal} {\bibinfo  {journal} {Rev. Mod. Phys.}\ }\textbf {\bibinfo {volume} {75}},\ \bibinfo {pages} {657} (\bibinfo {year} {2003})}\BibitemShut {NoStop}%
\bibitem [{\citenamefont {Yosida}(1958)}]{yosida_paramagnetic_1958}%
  \BibitemOpen
  \bibfield  {author} {\bibinfo {author} {\bibfnamefont {K.}~\bibnamefont {Yosida}},\ }\href {https://doi.org/10.1103/PhysRev.110.769} {\bibfield  {journal} {\bibinfo  {journal} {Phys. Rev.}\ }\textbf {\bibinfo {volume} {110}},\ \bibinfo {pages} {769} (\bibinfo {year} {1958})}\BibitemShut {NoStop}%
\bibitem [{\citenamefont {Deguchi}\ \emph {et~al.}(2004)\citenamefont {Deguchi}, \citenamefont {Mao}, \citenamefont {Yaguchi},\ and\ \citenamefont {Maeno}}]{deguchi_gap_2004}%
  \BibitemOpen
  \bibfield  {author} {\bibinfo {author} {\bibfnamefont {K.}~\bibnamefont {Deguchi}}, \bibinfo {author} {\bibfnamefont {Z.~Q.}\ \bibnamefont {Mao}}, \bibinfo {author} {\bibfnamefont {H.}~\bibnamefont {Yaguchi}},\ and\ \bibinfo {author} {\bibfnamefont {Y.}~\bibnamefont {Maeno}},\ }\href {https://doi.org/10.1103/PhysRevLett.92.047002} {\bibfield  {journal} {\bibinfo  {journal} {Phys. Rev. Lett.}\ }\textbf {\bibinfo {volume} {92}},\ \bibinfo {pages} {047002} (\bibinfo {year} {2004})}\BibitemShut {NoStop}%
\bibitem [{\citenamefont {Kittaka}\ \emph {et~al.}(2018)\citenamefont {Kittaka}, \citenamefont {Nakamura}, \citenamefont {Sakakibara}, \citenamefont {Kikugawa}, \citenamefont {Terashima}, \citenamefont {Uji}, \citenamefont {Sokolov}, \citenamefont {Mackenzie}, \citenamefont {Irie}, \citenamefont {Tsutsumi}, \citenamefont {Suzuki},\ and\ \citenamefont {Machida}}]{kittaka_searching_2018}%
  \BibitemOpen
  \bibfield  {author} {\bibinfo {author} {\bibfnamefont {S.}~\bibnamefont {Kittaka}}, \bibinfo {author} {\bibfnamefont {S.}~\bibnamefont {Nakamura}}, \bibinfo {author} {\bibfnamefont {T.}~\bibnamefont {Sakakibara}}, \bibinfo {author} {\bibfnamefont {N.}~\bibnamefont {Kikugawa}}, \bibinfo {author} {\bibfnamefont {T.}~\bibnamefont {Terashima}}, \bibinfo {author} {\bibfnamefont {S.}~\bibnamefont {Uji}}, \bibinfo {author} {\bibfnamefont {D.~A.}\ \bibnamefont {Sokolov}}, \bibinfo {author} {\bibfnamefont {A.~P.}\ \bibnamefont {Mackenzie}}, \bibinfo {author} {\bibfnamefont {K.}~\bibnamefont {Irie}}, \bibinfo {author} {\bibfnamefont {Y.}~\bibnamefont {Tsutsumi}}, \bibinfo {author} {\bibfnamefont {K.}~\bibnamefont {Suzuki}},\ and\ \bibinfo {author} {\bibfnamefont {K.}~\bibnamefont {Machida}},\ }\href {https://doi.org/10.7566/JPSJ.87.093703} {\bibfield  {journal} {\bibinfo  {journal} {J. Phys. Soc. Jpn.}\ }\textbf {\bibinfo {volume} {87}},\ \bibinfo {pages} {093703} (\bibinfo {year} {2018})}\BibitemShut {NoStop}%
\bibitem [{\citenamefont {Suh}\ \emph {et~al.}(2020)\citenamefont {Suh}, \citenamefont {Menke}, \citenamefont {Brydon}, \citenamefont {Timm}, \citenamefont {Ramires},\ and\ \citenamefont {Agterberg}}]{suh_stabilizing_2020}%
  \BibitemOpen
  \bibfield  {author} {\bibinfo {author} {\bibfnamefont {H.~G.}\ \bibnamefont {Suh}}, \bibinfo {author} {\bibfnamefont {H.}~\bibnamefont {Menke}}, \bibinfo {author} {\bibfnamefont {P.~M.~R.}\ \bibnamefont {Brydon}}, \bibinfo {author} {\bibfnamefont {C.}~\bibnamefont {Timm}}, \bibinfo {author} {\bibfnamefont {A.}~\bibnamefont {Ramires}},\ and\ \bibinfo {author} {\bibfnamefont {D.~F.}\ \bibnamefont {Agterberg}},\ }\href {https://doi.org/10.1103/PhysRevResearch.2.032023} {\bibfield  {journal} {\bibinfo  {journal} {Phys. Rev. Res.}\ }\textbf {\bibinfo {volume} {2}},\ \bibinfo {pages} {032023} (\bibinfo {year} {2020})}\BibitemShut {NoStop}%
\bibitem [{\citenamefont {Matsuda}\ and\ \citenamefont {Shimahara}(2007)}]{matsuda_fuldeferrelllarkinovchinnikov_2007}%
  \BibitemOpen
  \bibfield  {author} {\bibinfo {author} {\bibfnamefont {Y.}~\bibnamefont {Matsuda}}\ and\ \bibinfo {author} {\bibfnamefont {H.}~\bibnamefont {Shimahara}},\ }\href {https://doi.org/10.1143/JPSJ.76.051005} {\bibfield  {journal} {\bibinfo  {journal} {J. Phys. Soc. Jpn.}\ }\textbf {\bibinfo {volume} {76}},\ \bibinfo {pages} {051005} (\bibinfo {year} {2007})}\BibitemShut {NoStop}%
\bibitem [{\citenamefont {Kinjo}\ \emph {et~al.}(2022)\citenamefont {Kinjo}, \citenamefont {Manago}, \citenamefont {Kitagawa}, \citenamefont {Mao}, \citenamefont {Yonezawa}, \citenamefont {Maeno},\ and\ \citenamefont {Ishida}}]{kinjo_superconducting_2022}%
  \BibitemOpen
  \bibfield  {author} {\bibinfo {author} {\bibfnamefont {K.}~\bibnamefont {Kinjo}}, \bibinfo {author} {\bibfnamefont {M.}~\bibnamefont {Manago}}, \bibinfo {author} {\bibfnamefont {S.}~\bibnamefont {Kitagawa}}, \bibinfo {author} {\bibfnamefont {Z.~Q.}\ \bibnamefont {Mao}}, \bibinfo {author} {\bibfnamefont {S.}~\bibnamefont {Yonezawa}}, \bibinfo {author} {\bibfnamefont {Y.}~\bibnamefont {Maeno}},\ and\ \bibinfo {author} {\bibfnamefont {K.}~\bibnamefont {Ishida}},\ }\href {https://doi.org/10.1126/science.abb0332} {\bibfield  {journal} {\bibinfo  {journal} {Science}\ }\textbf {\bibinfo {volume} {376}},\ \bibinfo {pages} {397} (\bibinfo {year} {2022})}\BibitemShut {NoStop}%
\end{thebibliography}%
\bibliographystyle{apsrev4-2}

\end{document}